
\def\ignore#1{}
 

\newcount\sectnum
\newcount\subsectnum
\newcount\eqnumber

\global\eqnumber=1\sectnum=0


\def\lab{(\the\sectnum.\the\eqnumber)}



\def\show#1{#1}



\def\smskip{\vskip 5 pt}
\def\medskip{\vskip 10 pt}
\def\bigskip{\vskip 15 pt}
\def\pn{\par\noindent}
\def\br{\break}

\def\bl{\bigl} 
\def\br{\bigr} 
\def\lf{\left}
\def\ri{\right}

\def\ol#1{\overline{#1}}

\def\half{{\scriptstyle {1\over 2}}}

\def\Fscr{{\cal F}}

\def\a{\alpha}

\def\b{\beta}

\def\g{\gamma}

\def\p{\pi}

\def\e{\epsilon}

\def\o{\omega}

\def\re{\Re}
\def\rn{\Re^n}

\def\gr{\nabla}

\def\tl{\tilde}

\def\old#1{}
\def\leaderfill{\leaders\hbox to 1em{\hss.\hss}\hfill}


\parindent=2pc
\baselineskip=15pt
\vsize=8.7 true in
\voffset=0.125 true in
\parskip=3pt


\def\minprob#1#2#3{$$\eqalign{&\hbox{minimize\ \ }#1\cr &\hbox{subject to\ \
}#2\cr}\ifnum 0=#3{}\else\eqno(#3)\fi$$}        
     
\def\maxprob#1#2#3{$$\eqalign{&\hbox{maximize\ \ }#1\cr &\hbox{subject to\ \
}#2\cr}\ifnum 0=#3{}\else\eqno(#3)\fi$$}        
     
\def\aligntwo#1#2#3#4#5{$$\eqalign{#1&#2\cr #3&#4\cr}
\ifnum 0=#5{}\else\eqno(#5)\fi$$}
\def\alignthree#1#2#3#4#5#6#7{$$\eqalign{#1&#2\cr #3&#4\cr #5&#6\cr}
\ifnum 0=#7{}\else\eqno(#7)\fi$$}


\def\eqnum{\eqno{\hbox{(\the\sectnum.\the\eqnumber)}\global\advance\eqnumber
by1}}

\def\eqnu{\eqno{\hbox{(\the\sectnum.\the\eqnumber)}\global\advance\eqnumber
by1}}

\newcount\examplnumber
\def\examplnum{\global\advance\examplnumber by1}

\newcount\figrnumber
\def\figrnum{\global\advance\figrnumber by1}

\newcount\propnumber
\def\propnum{\global\advance\propnumber by1}

\newcount\defnumber
\def\defnum{\global\advance\defnumber by1}

\newcount\lemmanumber
\def\lemmanum{\global\advance\lemmanumber by1}

\newcount\assumptionnumber
\def\assumptionnum{\global\advance\assumptionnumber by1}

\def\exampl{\the\sectnum.\the\examplnumber}
\def\figr{\the\sectnum.\the\figrnumber}
\def\propn{\the\sectnum.\the\propnumber}
\def\defn{\the\sectnum.\the\defnumber}
\def\lemman{\the\sectnum.\the\lemmanumber}
\def\assumptionn{\the\sectnum.\the\assumptionnumber}

\def\section#1{\goodbreak\vskip 3pc plus 6pt minus 3pt\leftskip=-2pc
   \global\advance\sectnum by 1\eqnumber=1
\global\examplnumber=1\figrnumber=1\propnumber=1\defnumber=1\lemmanumber=1\assumptionnumber=1%
   \line{\hfuzz=1pc{\hbox to 3pc{\bf 
   \vtop{\hfuzz=1pc\hsize=38pc\hyphenpenalty=10000\noindent\uppercase{\the\sectnum.\quad #1}}\hss}}
			\hfill}
			\leftskip=0pc\nobreak\tenf
			\vskip 1pc plus 4pt minus 2pt\noindent\ignorespaces}



\def\sect#1{\noindent\leftskip=-2pc\tenf
   \goodbreak\vskip 1pc plus 4pt minus 2pt
                \global\advance\subsectnum by 1\eqnumber=1
   \line{\hfuzz=1pc{\hbox to 3pc{\bf 
   \vtop{\hfuzz=1pc\hsize=38pc\hyphenpenalty=10000\noindent\uppercase{{\bf #1}}}\hss}}
                        \hfill}
   \leftskip=0pc\nobreak\tenf
                        \vskip 1pc plus 4pt minus 2pt\nobreak\noindent\ignorespaces}

\def\subsection#1{\noindent\leftskip=0pc\tenf
   \goodbreak\vskip 1pc plus 4pt minus 2pt
   \line{\hfuzz=1pc{\hbox to 3pc{\bf 
   \vtop{\hfuzz=1pc\hsize=38pc\hyphenpenalty=10000\noindent{\bf #1}}\hss}}
                        \hfill}
   \leftskip=0pc\nobreak\tenf
                        \vskip 1pc plus 4pt minus 2pt\nobreak\noindent\ignorespaces}
\def\subsubsection#1{\goodbreak\vskip 1pc plus 4pt minus 2pt
   \hfuzz=3pc\leftskip=0pc\noindent\tenit #1 \nobreak\tenf\vskip 6pt plus 1pt
                                minus 1pt\nobreak\ignorespaces\leftskip=0pc}
%

\def\beginexample#1{\noindent\goodbreak\vskip 6pt plus 1pt minus 1pt
\noindent
  \hbox {\bf Example #1\hss}
  \nobreak\vskip 4pt plus 1pt minus 1pt \nobreak\noindent\ninef
  \global\advance
                \leftskip by\parindent\pn}
\def\endexample{\vskip 12pt\tenf\par
  \global\advance\leftskip by -\parindent
  }

\def\beginexercise#1{\noindent\goodbreak\vskip 6pt plus 1pt minus 1pt \noindent\global\normalbaselineskip=12pt
  \hbox {\bf Exercise #1\hss}
  \nobreak\vskip 4pt plus 1pt minus 1pt 
  \nobreak\noindent\ninef\global\advance\leftskip
                        by\parindent\pn}
\def\endexercise{\vskip 12pt\tenf\par
  \global\advance\leftskip by -\parindent
  }

\def\beginsection#1{\noindent\goodbreak\vskip 6pt plus 1pt minus 1pt \noindent\global\normalbaselineskip=12pt
  \hbox {\it #1\hss}
  \vskip 0.1pt plus 1pt minus 1pt \nobreak\noindent\ninef\global\advance
                \leftskip by\parindent\noindent\pn}
\def\endsection{\vskip 12pt\tenf\par
  \global\advance\leftskip by -\parindent
}

%


\def\section#1{\goodbreak\vskip 3pc plus 6pt minus 3pt\leftskip=-2pc
   \global\advance\sectnum by 1\eqnumber=1
\global\examplnumber=1\figrnumber=1\propnumber=1\defnumber=1\lemmanumber=1\assumptionnumber=1\subsectnum=0%
   \line{\hfuzz=1pc{\hbox to 3pc{\bf 
   \vtop{\hfuzz=1pc\hsize=38pc\hyphenpenalty=10000\noindent\uppercase{\the\sectnum.\quad #1}}\hss}}
			\hfill}
			\leftskip=0pc\nobreak\tenf
			\vskip 1pc plus 4pt minus 2pt\noindent\ignorespaces}

\def\subsection#1{\noindent\leftskip=0pc\tenf
   \goodbreak\vskip 1pc plus 4pt minus 2pt
               \global\advance\subsectnum by 1
   \line{\hfuzz=1pc{\hbox to 3pc{\bf \the\sectnum.\the\subsectnum\ \ \
   \vtop{\hfuzz=1pc\hsize=38pc\hyphenpenalty=10000\noindent{\bf #1}}\hss}}
                        \hfill}
   \leftskip=0pc\nobreak\tenf
                        \vskip 1pc plus 4pt minus 2pt\nobreak\noindent\ignorespaces}

\def\subsubsection#1{\goodbreak\vskip 1pc plus 4pt minus 2pt
   \hfuzz=3pc\leftskip=0pc\noindent{\bf #1} \nobreak\vskip 6pt plus 1pt
                                minus 1pt\nobreak\ignorespaces\leftskip=0pc}



\def\proposition#1{\smskip\pn{\bf Proposition #1}\quad}
\def\proof{\smskip\pn{\bf Proof:}\quad} 
 \def\assumption#1{\smskip\pn{\bf Assumption #1}\quad}

 \def\qed{\quad{\bf
Q.E.D.} \par\bigskip}
\def\ref{\smskip\pn}

\def\chapter#1#2{{\bf \centerline{\helbigbig
{#1}}}\bigskip\bigskip{\bf \centerline{\helbigbig
{#2}}}\bigskip\bigskip} 



\def\longpapertitle#1#2#3{{\bf \centerline{\helbigb
{#1}}}\bigskip{\bf \centerline{\helbigb
{#2}}}\bigskip\bigskip{\centerline{
by}}\bigskip{\bf \centerline{
{#3}}}\bigskip\bigskip} 


\def\nitem#1{\smskip\item{#1}}

\newcount\alphanum
\newcount\romnum

\def\alphaenumerate{\ifcase\alphanum \or (a)\or (b)\or (c)\or (d)\or (e)\or
(f)\or (g)\or (h)\or (i)\or (j)\or (k)\fi}
\def\romenumerate{\ifcase\romnum \or (i)\or (ii)\or (iii)\or (iv)\or (v)\or
(vi)\or (vii)\or (viii)\or (ix)\or (x)\or (xi)\fi}

\def\alist{\begingroup\vskip10pt\alphanum=1
\parskip=2pt\parindent=0pt \leftskip=3pc
\everypar{\llap{{\rm\alphaenumerate\hskip1em}}\advance\alphanum by1}}

\def\nolist{\begingroup\vskip10pt\alphanum=0
\parskip=2pt\parindent=0pt \leftskip=3pc
\everypar{\llap{\global\advance\alphanum by1(\the\alphanum)\hskip1em}}}

\def\romlist{\begingroup\vskip10pt\romnum=1
\parskip=2pt\parindent=0pt \leftskip=5pc
\everypar{\llap{{\rm\romenumerate\hskip1em}}\advance\romnum by1}}



\long\def\fig#1#2#3{\vbox{\vskip1pc\vskip#1
\prevdepth=12pt \baselineskip=12pt
\vskip1pc
\hbox to\hsize{\hfill\vtop{\hsize=25pc\noindent{\eightbf Figure #2\ }
{\eightpoint#3}}\hfill}}}

\long\def\widefig#1#2#3{\vbox{\vskip1pc\vskip#1
\prevdepth=12pt \baselineskip=12pt
\vskip1pc
\hbox to\hsize{\hfill\vtop{\hsize=28pc\noindent{\eightbf Figure #2\ }
{\eightpoint#3}}\hfill}}}

\long\def\table#1#2{\vbox{\vskip0.5pc
\prevdepth=12pt \baselineskip=12pt
\hbox to\hsize{\hfill\vtop{\hsize=25pc\noindent{\eightbf Table #1\ }
{\eightpoint#2}}\hfill}}}

 
\def\rightheadline#1{\headline{\tenrm\hfil #1}}


\long\def\leftfig#1#2{\vbox{\smskip\hsize=220pt
\vtop{{\noindent {\bf #1}}}
\smskip
\noindent
\vbox{{\noindent #2}}
}}

\long\def\rightfig#1#2#3{\vbox{\smskip\vskip#1
\prevdepth=12pt \baselineskip=12pt
\hsize=210pt
\smskip
\vbox{\noindent{\eightbold #2}
\hskip1em{\eightpoint#3}}
}}

\long\def\concept#1#2#3#4#5{\bigskip\hrule
\vbox{\hbox{\leftfig{#1}{#2} \hskip3em
\rightfig{#3}{#4}{#5}} \smskip}
\hrule\bigskip}


\long\def\bconcept#1#2#3#4#5#6#7{
\vbox{
\hbox to \hsize{\vtop{\par #1}}
\concept{#2}{#3}{#4}{#5}{#6}
\hbox to \hsize{\vtop{\par #7}}
\smskip}
}




\def\boxit#1{\vbox{\hrule\hbox{\vrule\kern3pt
                                \vbox{\kern3pt#1\kern3pt}\kern3pt\vrule}\hrule}}
\def\centerboxit#1{$$\vbox{\hrule\hbox{\vrule\kern3pt
                                \vbox{\kern3pt#1\kern3pt}\kern3pt\vrule}\hrule}$$}

\long\def\boxtext#1#2{$$\boxit{\vbox{\hsize #1\noindent\strut #2\strut}}$$}

%
%
%

\def\picture #1 by #2 (#3){
  \vbox to #2{
    \hrule width #1 height 0pt depth 0pt
    \vfill
    \special{picture #3} 
    }
  }

\def\scaledpicture #1 by #2 (#3 scaled #4){{
  \dimen0=#1 \dimen1=#2
  \divide\dimen0 by 1000 \multiply\dimen0 by #4
  \divide\dimen1 by 1000 \multiply\dimen1 by #4
  \picture \dimen0 by \dimen1 (#3 scaled #4)}
  }

%
%

\long\def\captfig#1#2#3#4#5{\vbox{\vskip1pc
\hbox to\hsize{\hfill{\picture #1 by #2 (#3)}\hfill}
\prevdepth=9pt \baselineskip=9pt
\vskip1pc
\hbox to\hsize{\hfill\vtop{\hsize=24pc\noindent{\eightbold Figure #4}
\hskip1em{\eightpoint#5}}\hfill}}}

%
%
%

\def\illustration #1 by #2 (#3){
  \vskip#2\hskip#1\special{illustration #3} 
    }

\def\scaledillustration #1 by #2 (#3 scaled #4){{
  \dimen0=#1 \dimen1=#2
  \divide\dimen0 by 1000 \multiply\dimen0 by #4
  \divide\dimen1 by 1000 \multiply\dimen1 by #4
  \illustration \dimen0 by \dimen1 (#3 scaled #4)}
  }


\newbox\graybox
\newdimen\xgrayspace
\newdimen\ygrayspace
%
%
%
%
%
%
%
%
%

\def\Textshade#1#2#3#4#5#6{%
    \xgrayspace=#4pt%
    \ygrayspace=#4pt%
    \def\grayshade{#3}%
    \def\linewidth{#5}%
    \def\theradius{#6}%
    \setbox\graybox=\hbox{\surroundboxa{#2}}%
    \hbox{%
    \hbox to 0pt{%
    \PScommands
}%
    \box\graybox}}%
%
%
\long%

\long%
\def\Parashade#1#2#3#4#5#6#7{%
    \xgrayspace=#4pt%
    \ygrayspace=#4pt%
    \def\grayshade{#3}%
    \def\linewidth{#5}%
    \def\theradius{#6}%
    \def\thevskip{#7pt}%
    \setbox\graybox=\hbox{\surroundboxb{#2}}%
    \vskip\thevskip%
    \hbox{%
    \hbox to 0pt{%
    \PScommands
}%
     \box\graybox}%
     \vskip\thevskip%
}%
%
%
%
\long\def\surroundboxa#1{\leavevmode\hbox{\vtop{%
\vbox{\kern\ygrayspace%
\hbox{\kern\xgrayspace#1%
      \kern\xgrayspace}}\kern\ygrayspace}}}
%
%
\long\def\surroundboxb#1{\leavevmode\hbox{\vtop{%
\vbox{\kern\ygrayspace%
\hbox{\kern\xgrayspace\vbox{\advance\hsize-2\xgrayspace#1}%
      \kern\xgrayspace}}\kern\ygrayspace}}}
%
%
%
\long\def\PScommands{%
\special{rawpostscript
/sharpbox{%
           newpath
           xmin ymin moveto
           xmin ymax lineto
           xmax ymax lineto
           xmax ymin lineto
           xmin ymin lineto
           closepath 
          }bind def
}%
\special{rawpostscript
/sharpboxnb{%
           newpath
           xmin ymin moveto
           xmin ymax lineto
           xmax ymax lineto
           xmax ymin lineto
          }bind def
}%
\special{rawpostscript
/sharpboxnt{%
           newpath
           xmin ymax moveto
           xmin ymin lineto
           xmax ymin lineto
           xmax ymax lineto
          }bind def
}%
\special{rawpostscript
/roundbox{%
           newpath
           xmin radius add ymin moveto
           xmax ymin xmax ymax radius arcto
           xmax ymax xmin ymax radius arcto
           xmin ymax xmin ymin radius arcto
           xmin ymin xmax ymin radius arcto 16 {pop} repeat
           closepath
          }bind def
}%
\special{rawpostscript
/sharpcorners{%
               sharpbox gsave grayshade setgray fill grestore 
               linewidth setlinewidth stroke
              }bind def
}%
\special{rawpostscript
/sharpcornersnt{%
               sharpboxnt gsave grayshade setgray fill grestore 
               linewidth setlinewidth stroke
              }bind def
}%
\special{rawpostscript
/sharpcornersnb{%
               sharpboxnb gsave grayshade setgray fill grestore 
               linewidth setlinewidth stroke
              }bind def
}%
\special{rawpostscript
/roundcorners{%
               roundbox gsave grayshade setgray fill grestore 
               linewidth setlinewidth stroke
              }bind def
}%
\special{rawpostscript
/plainbox{%
           sharpbox grayshade setgray fill 
          }bind def
}%
%
\special{rawpostscript
/roundnoframe{%
               roundbox grayshade setgray fill 
              }bind def
}%
\special{rawpostscript
/sharpnoframe{%
               sharpbox grayshade setgray fill 
              }bind def
}%
}%
%
%

\def\pshade#1{\Parashade{sharpcorners}{#1}{0.95}{10}{0.5}{10}{10}}


\def\boxit#1{\vbox{\hrule\hbox{\vrule\kern3pt
                                \vbox{\kern3pt#1\kern3pt}\kern3pt\vrule}\hrule}}

\def\boxitnb#1{\vbox{\hrule\hbox{\vrule\kern3pt
                                \vbox{\kern3pt#1\kern3pt}\kern3pt\vrule}}}

\def\boxitnt#1{\vbox{\hbox{\vrule\kern3pt
                                \vbox{\kern3pt#1\kern3pt}\kern3pt\vrule}\hrule}}

\long\def\boxtext#1#2{$$\boxit{\vbox{\hsize #1\noindent\strut #2\strut}}$$}
\long\def\boxtextnb#1#2{$$\boxitnb{\vbox{\hsize #1\noindent\strut #2\strut}}$$}
\long\def\boxtextnt#1#2{$$\boxitnt{\vbox{\hsize #1\noindent\strut #2\strut}}$$}

\def\texshopbox#1{\boxtext{462pt}{\vskip-1.5pc\pshade{\vskip-1.0pc#1\vskip-2.0pc}}}
\def\texshopboxnt#1{\boxtextnt{462pt}{\vskip-1.5pc\pshade{\vskip-1.0pc#1\vskip-2.0pc}}}
\def\texshopboxnb#1{\boxtextnb{462pt}{\vskip-1.5pc\pshade{\vskip-1.0pc#1\vskip-2.0pc}}}


%
%
%
%
%
%
%
%
\font\helbigbig=cmr10 scaled 2500%
\font\helbigb=cmbx10 scaled 1500%
\font\eightbold=cmbx8%

\def\tenf{\hel}%
\def\tenit{\heli}%
\def\ninef{\ninehel}%
\def\nineit{\nineheli}%
%
%


\font\tenrm=cmr10%
\font\teni=cmmi10%
\font\tensy=cmsy10%
\font\tenbf=cmbx10%
\font\tentt=cmtt10%
\font\tenit=cmti10%
\font\tensl=cmsl10%

\def\tenpoint{\def\rm{\fam0\tenrm}%
\textfont0=\tenrm%
\textfont1=\teni%
\textfont2=\tensy%
\textfont\itfam=\tenit%
\textfont\slfam=\tensl%
\textfont\ttfam=\tentt%
\textfont\bffam=\tenbf%
\scriptfont0=\sevenrm%
\scriptfont1=\seveni%
\scriptfont2=\sevensy%
\scriptscriptfont0=\sixrm%
\scriptscriptfont1=\sixi%
\scriptscriptfont2=\sixsy%
\def\it{\fam\itfam\tenit}%
\def\tt{\fam\ttfam\tentt}%
\def\sl{\fam\slfam\tensl}%
\scriptfont\bffam=\sevenbf%
\scriptscriptfont\bffam=\sixbf%
\def\bf{\fam\bffam\tenbf}%
\normalbaselineskip=18pt%
\normalbaselines\rm}%

\font\ninerm=cmr9%
\font\ninebf=cmbx9%
\font\nineit=cmti9%
\font\ninesy=cmsy9%
\font\ninei=cmmi9%
\font\ninett=cmtt9%
\font\ninesl=cmsl9%

\def\ninepoint{\def\rm{\fam0\ninerm}%
\textfont0=\ninerm%
\textfont1=\ninei%
\textfont2=\ninesy%
\textfont\itfam=\nineit%
\textfont\slfam=\ninesl%
\textfont\ttfam=\ninett%
\textfont\bffam=\ninebf%
\scriptfont0=\sixrm%
\scriptfont1=\sixi%
\scriptfont2=\sixsy%
\def\it{\fam\itfam\nineit}%
\def\tt{\fam\ttfam\ninett}%
\def\sl{\fam\slfam\ninesl}%
\scriptfont\bffam=\sixbf%
\scriptscriptfont\bffam=\fivebf%
\def\bf{\fam\bffam\ninebf}%
\normalbaselineskip=16pt%
\normalbaselines\rm}%

\font\eightrm=cmr8%
\font\eighti=cmmi8%
\font\eightsy=cmsy8%
\font\eightbf=cmbx8%
\font\eighttt=cmtt8%
\font\eightit=cmti8%
\font\eightsl=cmsl8%

\def\eightpoint{\def\rm{\fam0\eightrm}%
\textfont0=\eightrm%
\textfont1=\eighti%
\textfont2=\eightsy%
\textfont\itfam=\eightit%
\textfont\slfam=\eightsl%
\textfont\ttfam=\eighttt%
\textfont\bffam=\eightbf%
\scriptfont0=\sixrm%
\scriptfont1=\sixi%
\scriptfont2=\sixsy%
\scriptscriptfont0=\fiverm%
\scriptscriptfont1=\fivei%
\scriptscriptfont2=\fivesy%
\def\it{\fam\itfam\eightit}%
\def\tt{\fam\ttfam\eighttt}%
\def\sl{\fam\slfam\eightsl}%
\scriptscriptfont\bffam=\fivebf%
\def\bf{\fam\bffam\eightbf}%
\normalbaselineskip=14pt%
\normalbaselines\rm}%

\font\sevenrm=cmr7%
\font\seveni=cmmi7%
\font\sevensy=cmsy7%
\font\sevenbf=cmbx7%

\font\sixrm=cmr6%
\font\sixi=cmmi6%
\font\sixsy=cmsy6%
\font\sixbf=cmbx6%

\fontdimen13\tensy=2.6pt%
\fontdimen14\tensy=2.6pt%
\fontdimen15\tensy=2.6pt%
\fontdimen16\tensy=1.2pt%
\fontdimen17\tensy=1.2pt%
\fontdimen18\tensy=1.2pt%

\def\tenf{\tenpoint}%
\def\ninef{\ninepoint}%
%




\def\texshopbox#1{\boxtext{462pt}{\vskip-1.5pc\pshade{\vskip-1.0pc#1\vskip-2.0pc}}}
\def\texshopboxnt#1{\boxtextnt{462pt}{\vskip-1.5pc\pshade{\vskip-1.0pc#1\vskip-2.0pc}}}
\def\texshopboxnb#1{\boxtextnb{462pt}{\vskip-1.5pc\pshade{\vskip-1.0pc#1\vskip-2.0pc}}}


\input miniltx

\ifx\pdfoutput\undefined
  \def\Gin@driver{dvips.def} 
\else
  \def\Gin@driver{pdftex.def} 
\fi

\input graphicx.sty
\resetatcatcode

\long\def\fig#1#2#3{\vbox{\vskip1pc\vskip#1
\prevdepth=12pt \baselineskip=12pt
\vskip1pc
\hbox to\hsize{\hskip3pc\hfill\vtop{\hsize=35pc\noindent{\eightbf Figure #2\ }
{\eightpoint#3}}\hfill}}}

\def\show#1{}

\def\frac#1#2{{#1\over #2}}

\rightheadline{\botmark}

\pageno=1

\rightheadline{\botmark}

\pn {\bf August 2010 (revised December 2010)} \hfill{\bf  Report LIDS - 2848}
\bigskip \bigskip\bigskip

\bigskip\bigskip

\def\longpapertitle#1#2#3{{\bf \centerline{\helbigb
{#1}}}\medskip{\bf \centerline{\helbigb
{#2}}}\medskip{\centerline{
by}}\medskip{\bf \centerline{
{#3}}}\bigskip}

\def\longpapertitle#1#2#3{{\bf \centerline{\helbigb
{#1}}}\medskip{\bf \centerline{\helbigb
{#2}}}\medskip{\centerline{
}}\medskip{\bf \centerline{
{#3}}}\bigskip}

\vskip-3pc
\longpapertitle{Incremental Gradient, Subgradient, and Proximal Methods for}{Convex Optimization: A Survey\footnote{$\ ^1$}
{\ninepoint  This is an extended version of similarly titled papers that appear in Math.\ Programming Journal, 2011, Vol.\ 129, pp.\ 163-195, and the edited volume Optimization for Machine Learning (S.\ Sra, S.\ Nowozin, and S.\ Wright, Eds.), MIT Press, 2012. This version corrects two flaws of the Dec.\ 2010 original survey: in the statements and proofs of Props. 3.1 and 5.2. Both corrections are described by footnotes preceding the propositions. A supplementary survey, dealing with aggregated incremental gradient, proximal, and augmented Lagrangian methods is: Bertsekas, D.\ P., 2015.\  ``Incremental Aggregated Proximal and Augmented Lagrangian Algorithms," Lab.\ for Information and Decision Systems Report LIDS-P-3176, MIT, September 2015.}}
{Dimitri P.\ Bertsekas
\footnote{$^2$}
{\ninepoint  The author is with the Dept.\ of Electr.\ Engineering and
Comp.\ Science, M.I.T., Cambridge, Mass., 02139. His research was supported by the AFOSR under Grant FA9550-10-1-0412.  Thanks are due to Huizhen (Janey) Yu for extensive helpful discussions and suggestions. Comments by Angelia Nedi\'c and Ben Recht are also appreciated.}
}

\centerline{\bf Abstract}
We survey incremental methods for minimizing a sum $\sum_{i=1}^mf_i(x)$ consisting of a large number of convex component functions $f_i$. Our methods consist of iterations applied to single components, and have proved very effective in practice. We introduce a unified algorithmic framework for a variety of such methods, some involving gradient and subgradient iterations, which are known, and some involving combinations of subgradient and proximal methods, which are new and offer greater flexibility in exploiting the special structure of $f_i$. We provide an analysis of the convergence and rate of convergence properties of these methods, including the  advantages offered by randomization in the selection of components. We also survey applications in inference/machine learning, signal processing, and large-scale and distributed optimization.


\def\tl{\tilde}
\def\ol{\bar}

\def\dom{\hbox{dom}}

\def\old#1{}

\vskip  -4mm
\section{Introduction}
\vskip  -2mm

\pn We consider optimization problems with a cost function consisting of a large number of component functions, such as 
$$\eqalign{\hbox{\rm minimize}\quad &
\sum_{i=1}^mf_i(x)\cr
\hbox{\rm subject to\ \ }
&x\in X,\cr}\xdef\additive{\lab}\eqnum\show{twoo}$$
where $f_i:\rn\mapsto\re$, $i=1,\ldots,m$,  are real-valued functions, and $X$ is a closed convex set.\footnote\dag{\ninepoint Throughout the paper, we will operate within the $n$-dimensional space $\rn$ with the standard Euclidean norm, denoted $\|\cdot\|$. All vectors are considered column vectors and a prime denotes transposition, so $x'x=\|x\|^2$.  We will be using standard terminology of convex optimization, as given for example in textbooks such as Rockafellar's [Roc70], or the author's recent book [Ber09].} We focus on the case where the number of components $m$ is very large, and there is an incentive to use incremental methods that operate on a single component $f_i$ at each iteration, rather than on the entire cost function. If each incremental iteration tends to make reasonable progress in some ``average" sense, then depending on the value of $m$, an incremental method may significantly outperform (by orders of magnitude) its nonincremental counterpart, as experience has shown.

In this paper, we  survey the algorithmic properties of incremental methods in a unified framework, based on the author's recent work on incremental proximal methods [Ber10] (an early version appears in the supplementary algorithms chapter of the book [Ber09]). In this section, we first provide an overview of representative applications, and then we discuss three types of incremental methods: gradient, subgradient, and proximal. We unify these methods, into a combined method, which we use as a vehicle for analysis later.

\subsection{Some Examples of Additive Cost Problems}

\pn Additive cost problems of the form \additive\ arise in a variety of contexts. Let us provide a few examples where the incremental approach may have an advantage over alternatives.

 \xdef\examplemaxlike{\exampl}\examplnum\show{myexample}

\beginexample{\examplemaxlike: (Least Squares and Related Inference Problems)}An important context where cost functions of the form $\sum_{i=1}^mf_i(x)$ arise is inference/machine learning, where each term $f_i(x)$ corresponds to error between some data and the output of a parametric model, with $x$ being the vector of parameters. 
An example is {\it linear least squares} problems, where $f_i$ has quadratic structure, except for a regularization function. The latter function may be differentiable/quadratic, as in the classical regression problem 
\minprob{\sum_{i=1}^m(a_i'x-b_i)^2+\g \|x-\ol x\|^2}{x\in \rn,}{0}
where $\ol x$ is given, or nondifferentiable, as in the  {\it $\ell_1$-regularization problem}
\minprob{\sum_{i=1}^m(a_i'x-b_i)^2+\g \sum_{j=1}^n|x_j|}{x=(x_1,\ldots,x_n)\in \rn,}{0}
which will be discussed further in Section 5. 

A more general class of additive cost problems is {\it nonlinear least squares\/}. Here
$$f_i(x)=\big( h_i(x)\big)^2,$$
where $h_i(x)$ represents the difference between the $i$th of $m$ measurements from a physical system and the output of a parametric model whose parameter vector is $x$. Problems of nonlinear curve fitting and regression, as well as problems of training neural networks fall in this category, and they are typically nonconvex.

Another possibility is to use a nonquadratic function to penalize the error between some data and the output of the parametric model. For example in place of the squared error $(a_i'x-b_i)^2$, we may use
$$f_i(x)=\ell_i(a_i'x-b_i),$$
where $\ell_i$ is a convex function. This is a common approach in robust estimation and some support vector machine formulations.
 
Still another example is {\it maximum likelihood estimation\/}, where $f_i$ is of the form
$$f_i(x)=-\log P_Y(y_i;x),$$
and $y_1,\ldots,y_m$ represent values of independent samples of a random vector whose distribution $P_Y(\cdot;x)$ depends on an unknown parameter vector $x\in\rn$ that we wish to estimate. Related contexts include ``incomplete" data cases, where the expectation-maximization (EM) approach is used.
\endexample

The following four examples deal with broadly applicable problem structures that  give rise to additive cost functions. 
 
 \xdef\exampledualopt{\exampl}\examplnum\show{myexample}

\beginexample{\exampledualopt: (Dual Optimization in Separable Problems)}Consider the problem
$$\eqalign{\hbox{\rm
maximize\ \ } &\sum_{i=1}^mc_i(y_i)\cr \hbox{\rm subject to\ }
&\sum_{i=1}^mg_{i}(y_i)\ge0,\quad y_i\in Y_i,\ \ i=1,\ldots,m,}$$
where  $c_{i}:\re^\ell\mapsto\re$ and $g_{i}:\re^\ell\mapsto\re^n$ are functions of a vector $y_i\in \re^\ell$, 
and $Y_i$ are given
sets of $\re^\ell$. Then by assigning a dual vector/multiplier $x\in\re^n$
to the $n$-dimensional constraint function, we obtain the dual problem
$$\eqalign{\hbox{\rm minimize}\quad &\sum_{i=1}^nf_{i}(x)\cr
\hbox{\rm subject to\ \ }
&x\ge0,\cr}$$
where 
$$f_i(x)=\sup_{y_i\in Y_i}\lf\{c_i(y_i)+x'g_i(y_i)\ri\},$$
which has the additive form \additive.
Here $Y_i$ is not assumed convex, so integer programming and other discrete optimization problems are included. However, the dual cost function components $f_i$ are always convex, and their values and subgradients can often be conveniently computed, particularly when $y_i$ is a scalar or $Y_i$ is a finite set. 
\endexample

 \xdef\examplemanyc{\exampl}\examplnum\show{myexample}

\beginexample{\examplemanyc: (Problems with Many Constraints)}Problems of the form 
$$\eqalign{\hbox{\rm minimize}\quad &f(x)\cr
\hbox{\rm subject to\ \ }
&g_j(x)\le 0,\ \ j=1,\ldots,r,\ \ \ x\in X,\cr}\xdef\costfabt{\lab}\eqnum\show{oneo}$$
where the number $r$ of constraints is very large often arise in practice, either directly or via reformulation from other problems. They can be handled in a variety of ways. One possibility is to adopt a penalty function approach, and  replace problem \costfabt\ with
$$\eqalign{\hbox{\rm minimize}\quad &f(x)+c\sum_{j=1}^rP\big(g_j(x)\big)\cr
\hbox{\rm subject to\ \ }
&x\in X,\cr}\xdef\costfac{\lab}\eqnum\show{oneo}$$
where $P(\cdot)$ is a scalar penalty function satisfying $P(t)=0$ if $t\le0$, and $P(t)>0$ if $t>0$, and $c$ is a positive penalty parameter. For example, one may use the quadratic penalty 
$P(t)=\bl(\max\{0,t\}\br)^2,$
or the nondifferentiable penalty
 $P(t)= \max\{0,t\}.$
In the latter case, it can be shown that the optimal solutions of problems \costfabt\ and \costfac\ coincide when $c$ is sufficiently large (see for example [BNO03], Section 7.3, for the case where $f$ is convex). 
The cost function of the penalized problem \costfac\ is of the additive form \additive.

Set constraints of the form $x\in\cap_{i=1}^mX_i$, where $X_i$ are closed sets, can also be handled by penalties in a way that gives rise to additive cost functions (a simpler but important special case where such constraints arise is the problem of finding a common point within the sets $X_i$, $i=1,\ldots,m$; see Section 5.2).  In particular,  under relatively mild conditions, problem \costfabt\ with $X=\cap_{i=1}^mX_i$ is equivalent to the unconstrained minimization of 
$$f(x)+c\sum_{j=1}^rP\big(g_j(x)\big)+\g\sum_{i=1}^m\hbox{dist}(x;X_i),$$
where $\hbox{dist}(x;X_i)=\min_{y\in X_i}\|y-x\|$ and $\g$ is a sufficiently large penalty parameter. We discuss this possibility in Section 5.2.
\endexample

 \xdef\examplestochpr{\exampl}\examplnum\show{myexample}

\beginexample{\examplestochpr: (Minimization of an Expected Value - Stochastic Programming)}Consider the minimization of an expected value 
$$\eqalign{\hbox{\rm minimize}\quad &E\bl\{H(x,w)\br\}\cr
\hbox{\rm subject to\ \ }
&x\in X,\cr}\xdef\expectedcost{\lab}\eqnum\show{oneo}$$
where $H$ is a function of $x$ and a random variable $w$ taking a finite but very large number of values $w_i$, $i=1,\ldots,m$, with corresponding probabilities $\p_i$. Here the cost function can be written as the sum of the $m$ functions $\p_i H(x,w_i)$.

An example is {\it stochastic programming\/}, a classical model of two-stage optimization under uncertainty, where a vector
$x\in X$ is selected at cost $C(x)$, a random event occurs that has $m$ possible outcomes $w_1,\ldots,w_m$, and then another
vector
$y$ is selected from some set $Y$ with knowledge of the outcome that occurred. Then the optimal decision problem is to specify a vector $y_i\in Y$ for each outcome $w_i$, and
to minimize over $x$ and $y_i$ the expected cost
$$C(x)+\sum_{i=1}^m\p_iG_i(y_i),$$
where $G_i(y_i)$ is the cost associated with the occurrence of $w_i$ and $\p_i$ is the
corresponding probability. This is a problem with an additive cost function.

Additive cost function problems also arise from problem \expectedcost\ in a different way, when the expected value $E\bl\{H(x,w)\br\}$ is approximated by an $m$-sample average
$$F(x)={1\over m}\sum_{i=1}^mH(x,w_i),$$
where $w_i$ are independent samples of the random variable $w$. The minimum of the sample average $f(x)$ is then taken as an approximation of the minimum of $E\bl\{H(x,w)\br\}$.

\endexample

\vskip-1pc

\xdef\examplemanyc{\exampl}\examplnum\show{myexample}

\beginexample{\examplemanyc: (Weber Problem in Location Theory)}A basic problem in location theory is to find a point
$x$ in the plane whose sum of weighted distances from a given set of points
$y_1,\ldots,y_m$ is minimized. Mathematically, the problem is
\minprob{\sum_{i=1}^mw_i\|x-y_i\|}{x\in\rn,}{0} where  $w_1,\ldots,w_m$ are given
positive scalars. This problem descends from the famous Fermat-Torricelli-Viviani problem (see [BMS99] for an account of the history). The algorithmic approaches of the present paper would be of potential interest when the number of points $m$ is large.  We refer  to Drezner and Hamacher [DrH04] for a survey of recent research, and to Beck and Teboulle [BeT10] for a discussion that is relevant to our context.
\endexample


The structure of the additive cost function \additive\ often facilitates the use of a distributed computing system that  is well-suited for the incremental approach. The following is an illustrative example.

\xdef\examplesensor{\exampl}\examplnum\show{myexample}

\beginexample{\examplesensor: (Distributed Incremental Optimization -- Sensor Networks)}Consider a network of $m$ sensors where data are collected and are used to
  solve some inference problem involving a parameter vector $x$. If $f_i(x)$
  represents an error penalty for the data collected by the $i$th sensor, the
  inference problem is of the form \additive. While it is possible to
  collect all the data at a fusion center where the problem will be solved in
  centralized manner, it may be preferable to adopt a distributed approach in
  order to save in data communication overhead and/or take advantage of
  parallelism in computation. In such an approach the current iterate $x_k$ is
  passed on from one sensor to another, with each sensor $i$ performing an
  incremental iteration involving just its local component function $f_i$, and
  the entire cost function need not be known at any one location. We refer to
  Blatt, Hero, and Gauchman [BHG08], and Rabbat and Nowak [RaN04], [RaN05] for
  further discussion. 
  
  The approach of computing incrementally the values and subgradients of the components $f_i$ in a distributed manner can be substantially extended to apply to general systems of asynchronous distributed computation, where the components are processed at the nodes of a computing network, and the results are suitably combined,  as discussed by
Nedi\'c, Bertsekas, and Borkar [NBB01]. The analysis here relies on ideas from distributed asynchronous gradient methods (both deterministic and stochastic), which were developed in the early 80s by the author and his coworkers [Ber83], [TBA86], [BeT89]), and have been experiencing a resurgence recently (see e.g., Nedi\'c and Ozdaglar [NeO09]).
\endexample

\vskip-1pc

\subsection{Incremental Gradient Methods - Differentiable Problems}

\pn Let us consider first the case where the components $f_i$ are differentiable (not necessarily convex). Then, we may use incremental gradient methods, which have the form
$$x_{k+1}=P_X\big(x_k-\a_k \gr f_{i_k}(x_k)\big),\xdef\incrgr{\lab}\eqnum\show{twoo}$$
where $\a_k$ is a positive stepsize, $P_X(\cdot)$ denotes projection on $X$, and $i_k$ is the index of the cost component that is iterated on.  Such methods have a long history, particularly for the unconstrained case ($X=\rn)$, starting with the
Widrow-Hoff least mean squares (LMS) method [WiH60] for positive semidefinite quadratic component functions (see e.g., [Luo91], [BeT96], Section 3.2.5, [Ber99], Section 1.5.2). They  have also been used extensively for the training of neural networks, a case of nonquadratic/nonconvex cost components, under the generic name ``backpropagation methods." There are several variants of these methods, which differ in the  stepsize selection scheme, and the order  in which components are taken up for iteration (it could be deterministic or randomized). They are
supported by convergence analyses under various conditions; see Luo [Luo91], Grippo [Gri93], [Gri00], Luo and Tseng [LuT94], Mangasarian and Solodov [MaS94],  Bertsekas [Ber97], Solodov [Sol98], Tseng [Tse98]. 

When comparing the incremental gradient method with its classical nonincremental gradient counterpart [$m=1$ and all components lumped into a single function $F(x)=\sum_{i=1}^mf_i(x)$], there are two complementary performance issues to consider:

\nitem{(a)} {\it Progress when far from convergence}. Here  the incremental method can be much faster. For an extreme case let $X=\rn$ (no constraints), and take $m$ very large and all components $f_i$ identical to each other. Then an  incremental iteration requires $m$ times less computation than a classical gradient iteration, but gives exactly the same result, when the stepsize is appropriately scaled to be $m$ times larger. While this is an extreme example, it reflects the essential mechanism by which incremental methods can be far superior: when the components $f_i$ are not too dissimilar, far from the minimum a single component gradient will point to ``more or less" the right direction [see also the discussion of [Ber97], and [Ber99] (Example 1.5.5 and Exercise 1.5.5)].

\nitem{(b)} {\it Progress when close to convergence}. Here the incremental method is generally inferior. As we will discuss shortly, it converges at a sublinear rate because it requires a diminishing stepsize $\a_k$, compared with the typically linear rate achieved with the classical gradient method when a small constant stepsize is used ($\a_k\equiv \a$). One may use a constant stepsize with the incremental method, and indeed this may be the preferred mode of implementation, but then the method typically oscillates in the neighborhood of a solution, with size of oscillation roughly proportional to $\a$, as examples and theoretical analysis show.
\smskip

To understand the convergence mechanism of incremental gradient methods, let us consider the case $X=\rn$, and assume  that the component functions $f_i$ are selected for iteration according to a cyclic order [i.e., $i_k=(k\hbox{ modulo }m)+1$], and let us assume that $\a_k$ is constant within a cycle (i.e., for all $\ell=0,1,\ldots$, $\a_{\ell m}=\a_{\ell m+1}=\cdots=\a_{\ell m+m-1}$). Then, viewing the iteration \incrgr\ in terms of cycles, we have for every $k$ that marks the beginning of a cycle ($i_k=1$),
$$x_{k+m}=x_{k}-\a_{k}\sum_{i=1}^m\gr f_i(x_{k+i-1})=x_{k}-\a_{k}\big(\gr F(x_k)+e_k\big), \xdef\graderr{\lab}\eqnum\show{target} $$
where $F$ is the cost function/sum of components,
$F(x)=\sum_{i=1}^mf_i(x),$
and $e_k$ is given by
$$e_k=\sum_{i=1}^m\big(\gr f_i(x_k)-\gr f_i(x_{k+i-1})\big),$$
and may be viewed as an error in the calculation of the gradient $\gr f(x_k)$. For Lipschitz continuous gradient functions $\gr f_i$, the error $e_k$ is proportional to $\a_k$, and this shows two fundamental properties of incremental gradient methods, which hold generally for the other incremental methods of this paper as well:

\nitem{(a)} A constant stepsize ($\a_k\equiv\a$) typically cannot guarantee convergence, since then the size of the gradient error $\|e_k\|$ is typically bounded away from 0. Instead  (in the case of differentiable components $f_i$) a peculiar form of convergence takes place for constant but sufficiently small $\a$, whereby the iterates within cycles converge but to different points within a sequence of $m$ points (i.e., the sequence of first points in the cycles converges to a different limit than the sequence of second points in the cycles, etc). This is true even in the most favorable case of a linear least squares problem  (see Luo [Luo91], or the textbook analysis of [Ber99], Section 1.5.1).

\nitem{(b)} A diminishing stepsize  [such as $\a_k=O(1/k)$] leads to diminishing error $e_k$, so (under the appropriate Lipschitz condition) it can result in convergence to a stationary point of $f$.
\smskip

A corollary of these properties is that the price for achieving convergence is the slow (sublinear) asymptotic rate of convergence associated with a diminishing stepsize, which compares unfavorably with the often linear rate of convergence associated with a constant stepsize and the nonincremental gradient method. However, in practical terms this argument does not tell the entire story, since the incremental gradient method often achieves in the early iterations a much faster convergence rate than its nonincremental counterpart. In practice, the 
incremental method is usually operated with a stepsize that is either constant or is gradually reduced up to a positive value, which is small enough so that the resulting asymptotic oscillation is of no essential concern. An alternative, is to use a constant stepsize throughout, but reduce over time the degree of incrementalism, so that ultimately the method becomes nonincremental and achieves a linear convergence rate (see [Ber97], [Sol98]).

Aside from extensions to nonidifferentiable cost problems, for $X=\rn$, there is an important variant of the incremental gradient method that involves extrapolation along the direction of the difference of the preceding two iterates:
$$x_{k+1}=x_k-\a_k \gr f_{i_k}(x_k)+\b (x_k-x_{k-1}),\xdef\heavy{\lab}\eqnum\show{twoo}$$
where $\b$ is a scalar in $[0,1)$ and $x_{-1}=x_0$ (see e.g., [MaS94], [Tse98], [Ber96], Section 3.2). This is sometimes called {\it incremental gradient method with momentum\/}. The nonincremental version of this method is the {\it heavy ball} method of Polyak [Pol64], which can be shown to have faster convergence rate than the corresponding gradient method (see [Pol87], Section 3.2.1). A nonincremental method of this type, but with variable and suitably chosen value of $\b$, has been proposed by Nesterov [Nes83], and has received a lot of attention recently because it has optimal iteration complexity properties under certain conditions (see Nesterov [Nes04], [Nes05], Lu, Monteiro, and Yuan [LMY08], Tseng [Tse08], Beck and  Teboulle [BeT09], [BeT10]). However, no incremental analogs of this method with favorable complexity properties are currently known.

Another variant of the incremental gradient method for the case $X=\rn$ has been proposed  by Blatt, Hero, and Gauchman [BHG08], which (after the first $m$ iterates are computed) has the form
$$x_{k+1}=x_k-\a\sum_{\ell=0}^{m-1}\gr f_{i_{k-\ell}}(x_{k-\ell})\xdef\bhgincrgr{\lab}\eqnum\show{twoo}$$
[for $k<m$, the summation should go up to $\ell=k$, and $\a$ should be replaced by a corresponding larger value, such as $\a_k=m\a/(k+1)$]. This method also computes the gradient incrementally, one component per iteration, but  in place of the single component gradient $\gr f_{i_k}(x_k)$ in Eq.\ \incrgr, it uses an approximation to the total cost gradient $\gr f(x_k)$, which is an aggregate of the component gradients computed in the past $m$ iterations. A cyclic order of component function selection [$i_k=(k\hbox{ modulo }m)+1$] is assumed in [BHG08], and 
a convergence analysis is given, including a linear convergence rate result for a sufficiently small constant stepsize $\a$ and quadratic component functions $f_i$. It is not clear how iterations \incrgr\ and \bhgincrgr\ compare in terms of rate of convergence, although the latter seems likely to make faster progress when close to convergence. Note that iteration \bhgincrgr\ bears similarity to the incremental gradient iteration with momentum \heavy\ where $\b\approx 1$. In particular, when $\a_k\equiv\a$, the sequence generated by Eq.\ \heavy\ satisfies
$$x_{k+1}=x_k-\a\sum_{\ell=0}^{k}\b^\ell\,\gr f_{i_{k-\ell}}(x_{k-\ell})\xdef\heavya{\lab}\eqnum\show{twoo}$$
[both iterations \bhgincrgr\ and \heavya\ involve different types of diminishing dependence on past gradient components].  There are no known analogs of iterations \heavy\ and \bhgincrgr\ for nondifferentiable cost problems. 

Among alternative incremental methods for differentiable cost problems, let us also mention  versions of the Gauss-Newton method for nonlinear least squares problems, based on the extended Kalman filter (Davidon [Dav76], Bertsekas [Ber96], and Moriyama, Yamashita, and Fukushima [MYF03]). They are mathematically equivalent to the ordinary Gauss-Newton method for linear least squares, which they solve exactly after a single pass through the component functions $f_i$, but they often perform much faster than the latter in the nonlinear case, particularly when $m$ is large.

Let us finally note that incremental gradient methods are also related to stochastic gradient methods, which aim to minimize an expected value $E\bl\{H(x,w)\br\}$ (cf.\ Example 1.2) by using the iteration
$$x_{k+1}=x_k-\a_k \gr H(x_k,w_k),$$
where $w_k$ is a sample of the random variable $w$. These methods also have a long history (see  Polyak and Tsypkin [PoT73], Ljung [Lju77], Kushner and Clark [KuC78], Tsitsiklis, Bertsekas, and Athans [TBA86], Polyak [Pol87], Bertsekas and Tsitsiklis [BeT89],  [BeT96], [BeT00], Gaivoronskii [Gai93], Pflug  [Pfl96], Kushner and Yin [KuY97],  Bottou [Bot05], Meyn [Mey07], Borkar [Bor08],  Nemirovski et.\ al\ [NJL09], Lee and Wright [LeW10]), and are strongly connected with stochastic approximation algorithms. The
main difference between stochastic and deterministic formulations is that the former involve sequential sampling of cost components from an infinite population under some
statistical assumptions, while in the latter the set of cost components is predetermined and finite. However, it is possible to view the incremental gradient method \incrgr, with a randomized selection of the component function $f_i$ (i.e., with $i_k$ chosen to be any one of the indexes $1,\ldots,m$, with equal probability $1/m$), as a stochastic gradient method (see [BeT96], Example 4.4, [BeT00], Section 5).

The stochastic formulation of incremental methods just discussed highlights an important application context where the component functions $f_i$ are not given a priori, but rather become known sequentially through some observation process. Then it often makes sense to use an incremental method to process the component functions as they become available, and to obtain approximate solutions as early as possible. In fact this may be essential in time-sensitive and possibly time-varying environments, where solutions are needed ``on-line." In such cases, one may hope than an adequate estimate of the optimal solution will be obtained, before all the functions $f_i$ are processed for the first time.

\xdef\suf{\tl \gr f}
\xdef\suh{\tl \gr h}
\xdef\suF{\tl \gr F}

\subsection{Incremental Subgradient Methods - Nondifferentiable Problems}
\vskip-2mm

\pn
We now discuss the case where the component functions $f_i$ are convex and nondifferentiable at some points, and consider incremental subgradient methods. These are similar to their gradient counterparts \incrgr\ except that an arbitrary subgradient $\suf_{i_k}(x_k)$ of the cost component $f_{i_k}$ is used in place of the gradient:\footnote{\dag}
{\ninepoint  In this paper, we use $\suf(x)$ to denote a subgradient of a convex function $f$ at a vector $x$, i.e, a vector such that $f(z)\ge f(x)+\suf(x)'(z-x)$ for all $x\in\rn$. The choice of $\suf(x)$ from within the set of all subgradients at $x$ [the subdifferential at $x$, denoted $\partial f(x)$] will be clear from the context. Note that if $f$ is real-valued, $\partial f(x)$ is nonempty and compact for all $x$. If $f$ is differentiable at $x$, $\partial f(x)$ consists of a single element, the gradient $\gr f(x)$.}
$$x_{k+1}=P_X\big(x_k-\a_k \suf_{i_k}(x_k)\big).\xdef\incrsubgr{\lab}\eqnum\show{twoo}$$
Such methods were first proposed in the general form \incrsubgr\ in the Soviet Union by Kibardin [Kib80], following the earlier paper by Litvakov [Lit66] (which considered convex/nondifferentiable extensions of linear least squares problems)  and other related subsequent proposals.\footnote{\ddag}
{\ninepoint Generally, in the 60s and 70s, algorithmic ideas relating to simple gradient methods with and without deterministic and stochastic errors were popular in the Soviet scientific community, partly due to an emphasis on stochastic iterative algorithms, such as pseudogradient and stochastic approximation; the works of Ermoliev, Polyak, and Tsypkin, to name a few of the principal contributors, are representative [Erm69], [PoT73], [Erm76], [Pol78], [Pol87]. By contrast the emphasis in the Western literature at the time was in more complex Newton-like and conjugate direction methods.} These works remained unnoticed in the Western literature, where incremental methods were reinvented often in different contexts and with different lines of analysis; see Solodov and Zavriev [SoZ98], Bertsekas [Ber99] (Section 6.3.2), Ben-Tal, Margalit, and Nemirovski [BMN01], Nedi\'c and Bertsekas [NeB00],  [NeB01], [NeB10], Nedi\'c, Bertsekas, and Borkar [NBB01], Kiwiel [Kiw04], Rabbat and Nowak [RaN04], [RaN05], Gaudioso, Giallombardo, and Miglionico [GGM06], Shalev-Shwartz et.\ al.\ [SSS07], Helou and De Pierro [HeD09], Johansson, Rabi, and Johansson [JRJ09], Predd, Kulkarni, and Poor [PKP09], and Ram, Nedi\'c, Veeravalli [RNV09], [RNV09], and Duchi, Hazan, and Singer [DHS10]. 

Incremental subgradient methods have convergence characteristics that are similar in many ways to their gradient counterparts, the most important similarity being the necessity for a diminishing stepsize $\a_k$ for convergence. The lines of analysis, however, tend to be different, since incremental gradient methods rely for convergence on arguments based on decrease of the cost function value, while incremental subgradient methods rely on arguments based on decrease of the iterates' distance to the optimal solution set. The line of analysis of the present paper is of the latter type, similar to 
earlier works of the author and his collaborators (see [NeB00],  [NeB01], [NBB01], and the textbook presentations in [Ber99], [BNO03]).  

Note two important ramifications of the lack of differentiability of the component functions $f_i$:

\nitem{(1)} Convexity of $f_i$ becomes essential, since the notion of subgradient is connected with convexity (subgra\-dient-like algorithms for  nondifferentiable/nonconvex problems have been suggested in the literature, but tend to be complicated and have not found much application thus far). 

\nitem{(2)} There is more reason to favor the incremental over the nonincremental methods, since (contrary to the differentiable case) nonincremental subgradient methods also require a diminishing stepsize for convergence, and typically achieve a sublinear rate of convergence. Thus the one theoretical advantage of the nonincremental gradient method discussed earlier is not shared by its subgradient counterpart.
\smskip

Let us finally mention that just as in the differentiable case, there is a substantial literature for stochastic versions of subgradient methods. In fact, as we will discuss in this paper, there is a potentially significant advantage in turning the method into a stochastic one by randomizing the order of selection of the components $f_i$ for iteration.

\subsection{Incremental Proximal Methods}

\pn
We now consider an extension of the incremental approach to proximal algorithms. The simplest one for problem \additive\ is of the form
$$x_{k+1}=\arg\min_{x\in X}\left\{f_{i_k}(x)+{1\over 2\a_k}\|x-x_k\|^2\right\},
\xdef\incrprox{\lab}\eqnum\show{twoo}$$
which relates to the proximal minimization algorithm (Martinet [Mar70], Rockafellar [Roc76]) in the same way that the incremental subgradient method \incrsubgr\ relates to the classical nonincremental subgradient method.\footnote{\dag}
{\ninepoint  In this paper we restrict attention to proximal methods with the quadratic regularization term $\|x-x_k\|^2$. Our approach is applicable in principle when a nonquadratic term is used instead in order to match the structure of the given problem. The discussion of such alternative algorithms is beyond our scope.}
Here $\{\a_k\}$ is a positive scalar sequence, and we will assume that each $f_i:\rn\mapsto\re$ is a convex function, and $X$ is a nonempty closed convex set. The motivation for this type of method, which was proposed only recently in [Ber10], is that with a favorable structure of the components, the proximal iteration \incrsubgr\ may be obtained in closed form or be relatively simple, in which case it may be preferable to a gradient or subgradient iteration. In this connection, we note that generally, proximal iterations are considered more stable than gradient iterations; for example in the nonincremental case, they converge essentially for any choice of $\a_k$, while this is not so for gradient methods.

Unfortunately, while some cost function components may be well suited for a proximal iteration, others may not be because the minimization \incrprox\ is inconvenient, and this leads us to consider combinations of gradient/subgradient and proximal iterations. In fact this has motivated in the past nonincremental combinations of gradient and proximal methods for minimizing the sum of two functions (or more generally, finding a zero of the sum of two nonlinear operators). These methods have a long history, dating to the splitting algorithms of Lions and Mercier [LiM79], Passty [Pas79], and Spingarn [Spi85], and have become popular recently (see Beck and Teboulle [BeT09], [BeT10], and the references they give to specialized algorithms, such as shrinkage/thresholding, cf.\ Section 5.1). Let us also note that splitting methods are related to alternating direction methods of multipliers (see Gabay and Mercier [GaM76], [Gab83], Bertsekas and Tsitsiklis [BeT89], Eckstein and Bertsekas [EcB92]), which are presently experiencing a revival as viable (nonincremental) methods for minimizing sums of component functions (see the survey by Boyd et.\ al.\ [BPC10], which contains extensive references to recent work and applications, and the complexity-oriented work of Goldfarb, Ma, and Scheinberg [GoM09], [GMS10]).

With similar motivation in mind, we adopt in this paper a unified algorithmic framework that includes incremental gradient, subgradient, and proximal methods, and their combinations, and serves to highlight their common structure and behavior. We focus on  problems of the form 
$$\eqalign{\hbox{\rm minimize}\quad &
F(x)\;	{\buildrel\rm def\over=}\;\sum_{i=1}^mF_i(x)\cr
\hbox{\rm subject to\ \ }
&x\in X,\cr}\xdef\combadditive{\lab}\eqnum\show{twoo}$$
where for all $i$,
$$F_i(x)=f_i(x)+h_i(x),\xdef\combadditivefh{\lab}\eqnum\show{twoo}$$
$f_i:\rn\mapsto\re$ and $h_i:\rn\mapsto\re$ are real-valued convex functions, and $X$ is a nonempty closed convex set.

 In Section 2, we consider several incremental algorithms that iterate on the components $f_i$ with a proximal iteration, and on the components $h_i$ with a subgradient iteration.
By choosing all the $f_i$ or all the $h_i$ to be identically zero, we obtain as special cases the subgradient and proximal iterations \incrsubgr\ and \incrprox, respectively. However, our methods offer greater flexibility, and may exploit the special structure of problems where the functions $f_i$ are suitable for a proximal iteration, while the components $h_i$ are not and thus may be preferably treated with a subgradient iteration.

 In Section  3, we discuss the convergence and rate of convergence properties of methods that use a cyclic rule for component selection, while in Section 4, we discuss the case of a randomized component selection rule. In summary, the convergence behavior of our incremental methods is similar to the one outlined earlier for the incremental subgradient method \incrsubgr. This includes convergence within a certain error bound for a constant stepsize, exact convergence to an optimal solution for an appropriately diminishing stepsize, and improved convergence rate/iteration complexity when randomization is used to select the cost component for iteration. In Section 5 we illustrate our methods for some example applications.

\vskip  -5mm
\section{Incremental Subgradient-Proximal Methods}
\vskip  -2mm

\xdef\propincrproxo{\propn}\propnum\show{myproposition}

\pn 
In this section, we consider problem \combadditive-\combadditivefh, and introduce several incremental algorithms that involve a combination of a proximal and a subgradient iteration. One of our algorithms has the form
$$z_k=\arg\min_{x\in X}\left\{f_{i_k}(x)+{1\over 2\a_k}\|x-x_k\|^2\right\},
\xdef\combincrprox{\lab}\eqnum\show{twoo}$$
 $$x_{k+1}=P_X\big(z_k-\a_k \suh_{i_k}(z_k)\big),\xdef\combincrsubgr{\lab}\eqnum\show{twoo}$$
where $\suh_{i_k}(z_k)$ is an arbitrary subgradient of $h_{i_k}$ at $z_k$. Note that the iteration is well-defined because  the minimum in Eq.\ \combincrprox\ is uniquely attained since $f_i$ is continuous and $\|x-x_k\|^2$ is real-valued, strictly convex, and coercive, while the subdifferential $\partial h_i(z_k)$ is nonempty since $h_i$ is real-valued.  Note also that by choosing all the $f_i$ or all the $h_i$ to be identically zero, we obtain as special cases the subgradient and proximal iterations \incrsubgr\ and \incrprox, respectively.

The iterations \combincrprox\ and \combincrsubgr\ maintain both sequences $\{z_k\}$ and  $\{x_k\}$ within the constraint set $X$, but it may be convenient to relax this constraint for either the proximal or the subgradient iteration, thereby requiring a potentially simpler computation. This leads to the algorithm
$$z_k=\arg\min_{x\in \rn}\left\{f_{i_k}(x)+{1\over 2\a_k}\|x-x_k\|^2\right\},
\xdef\combincrproxa{\lab}\eqnum\show{twoo}$$
 $$x_{k+1}=P_X\big(z_k-\a_k \suh_{i_k}(z_k)\big),\xdef\combincrsubgra{\lab}\eqnum\show{twoo}$$
where the restriction $x\in X$ has been omitted from the proximal iteration, and the algorithm
$$z_k=x_k-\a_k \suh_{i_k}(x_k),\xdef\combincrproxb{\lab}\eqnum\show{twoo}$$
 $$x_{k+1}=\arg\min_{x\in X}\left\{f_{i_k}(x)+{1\over 2\a_k}\|x-z_k\|^2\right\},\xdef\combincrsubgrb{\lab}\eqnum\show{twoo}$$
where the projection onto $X$ has been omitted from the subgradient iteration.  It is also possible to use different stepsize sequences in the proximal and subgradient iterations, but for notational simplicity we will not discuss this type of algorithm.

All of the incremental proximal algorithms given above are new to our knowledge, having first been proposed in the author's recent paper [Ber10] and the on-line chapter of the book [Ber09]. The closest connection to the existing proximal methods is the ``proximal gradient" method, which has been analyzed and discussed recently in the context of several machine learning applications by Beck and Teboulle [BeT09], [BeT10] (it can also be interpreted in terms of splitting algorithms [LiM79], [Pas79]). This method is nonincremental, applies to differentiable $h_i$, and contrary to subgradient and incremental methods, it does not require a diminishing stepsize for convergence to the optimum. In fact, the line of convergence analysis of Beck and Teboulle relies on the differentiability of $h_i$ and the nonincremental character of the proximal gradient method, and is thus different from ours. 

Part (a) of the following proposition is a key fact about incremental proximal iterations. It shows that they are closely related to incremental subgradient iterations, with the only difference being that the subgradient is evaluated at the end point of the iteration rather than at the start point.
Part (b) of the proposition provides an inequality that is well-known in the theory of proximal methods, and will be useful for our convergence analysis. In the following, we denote by $\hbox{ri}(S)$ the relative interior of a convex set $S$, and by $\dom(f)$ the effective domain $\big\{x\mid f(x)<\infty\big\}$ of a function $f:\rn\mapsto(-\infty,\infty]$.

\texshopbox{\proposition{\propincrproxo:} Let $X$ be a nonempty closed convex set, and let $f:\rn\mapsto(-\infty,\infty]$ be a closed proper convex function such that $\hbox{ri}(X)\cap\hbox{ri} \big(\dom(f)\big)\ne\emptyset$. For any $x_k\in\rn$ and $\a_k>0$, consider the proximal iteration 
$$x_{k+1}=\arg\min_{x\in X}\left\{f(x)+{1\over 2\a_k}\|x-x_k\|^2\right\}.\xdef\proxiter{\lab}\eqnum\show{twoo}$$
\nitem{(a)} The iteration can be written as
$$x_{k+1}=P_{X}\big(x_k- \a_k \suf(x_{k+1})\big),\qquad i=1,\ldots,m,\xdef\proxitereq{\lab}\eqnum\show{twoo}$$
where $\suf(x_{k+1})$ is some subgradient of $f$ at $x_{k+1}$.
\nitem{(b)} For all $y\in X$, we have
$$\eqalign{\|x_{k+1}-y\|^2&\le \|x_k-y\|^2-2\a_k \big(f(x_{k+1})-f(y)\big)-\|x_k-x_{k+1}\|^2\cr
&\le \|x_k-y\|^2-2\a_k \big(f(x_{k+1})-f(y)\big).\cr}\eqnum\show{twoo}$$
\old{
\nitem{(c)} If $x_k\in X$, we have
$$\|x_{k+1}-x_k\|^2\le 2\a_k \big(f(x_k)-f(x_{k+1})\big).\old{\eqnum\show{twoo}}$$
}
}

\proof (a) We use the formula for the subdifferential of the sum of the three functions $f$, $(1/2\a_k)\|x-x_k\|^2$, and the indicator function of $X$ (cf.\ Prop.\ 5.4.6 of [Ber09]), together with the condition that 0 should belong to this subdifferential at the optimum $x_{k+1}$. We obtain that
Eq.\ \proxiter\ holds if and only if 
$${1\over \a_k}(x_k-x_{k+1})\in \partial f(x_{k+1})+N_X(x_{k+1}),\xdef\subgradrel{\lab}\eqnum\show{twoo}$$
where $N_X(x_{k+1})$ is the normal cone of $X$ at $x_{k+1}$ [the set of vectors $y$ such that $y'(x-x_{k+1})\le0$ for all $x\in X$, and also the subdifferential of the indicator function of $X$ at $x_{k+1}$; see [Ber09], p.\ 185]. This is true if and only if
$$x_k-x_{k+1}-\a_k \suf(x_{k+1})\in N_X(x_{k+1}),$$
for some $\suf(x_{k+1})\in \partial f(x_{k+1})$, which in turn is true if and only if
Eq.\ \proxitereq\ holds, by the projection theorem.
\smskip
\pn (b) We have 
$$\|x_k-y\|^2=\|x_k-x_{k+1}+x_{k+1}-y\|^2=\|x_k-x_{k+1}\|^2-2(x_k-x_{k+1})'(y-x_{k+1})+\|x_{k+1}-y\|^2.\xdef\relone{\lab}\eqnum\show{twoo}$$
Also since from Eq.\ \subgradrel, ${1\over \a_k}(x_k-x_{k+1})$ is a subgradient at $x_{k+1}$ of the sum of $f$ and the indicator function of $X$, we have (using also the assumption $y\in X$)
$$f(x_{k+1})+{1\over \a_k}(x_k-x_{k+1})'(y-x_{k+1})\le f(y).$$
Combining this relation with Eq.\ \relone, the result follows.
\qed

Based on Prop.\ \propincrproxo(a), we see that all the preceding iterations can be written in an incremental subgradient format:

\nitem{(a)} Iteration \combincrprox-\combincrsubgr\ can be written as
$$z_k=P_X\big(x_k-\a_k \suf_{i_k}(z_k)\big),\qquad x_{k+1}=P_X\big(z_k-\a_k \suh_{i_k}(z_k)\big).\xdef\combincrsubgrt{\lab}\eqnum\show{twoo}$$
 \nitem{(b)} Iteration \combincrproxa-\combincrsubgra\ can be written as
$$z_k=x_k-\a_k \suf_{i_k}(z_k),\qquad x_{k+1}=P_X\big(z_k-\a_k \suh_{i_k}(z_k)\big).\xdef\combincrsubgrat{\lab}\eqnum\show{twoo}$$
 \nitem{(c)} Iteration \combincrproxb-\combincrsubgrb\ can be written as
$$z_k=x_k-\a_k \suh_{i_k}(x_k),\qquad x_{k+1}=P_X\big(z_k-\a_k \suf_{i_k}(x_{k+1})\big).\xdef\combincrsubgrbt{\lab}\eqnum\show{twoo}$$

\smskip
\pn 
Note that in all the preceding updates, the subgradient $\suh_{i_k}$ can be {\it any} vector in the subdifferential of $h_{i_k}$, while the subgradient $\suf_{i_k}$ must be a {\it specific} vector in the subdifferential of $f_{i_k}$, specified according to Prop.\ \propincrproxo(a). 
Note also that iteration \combincrsubgrat\ can be written as
$$x_{k+1}=P_X\big(x_k- \a_k\suF_{i_k}(z_k)\big),$$
and resembles the incremental subgradient method for minimizing over $X$ the cost 
$F(x)=\sum_{i=1}^mF_i(x)$ [cf.\ Eq.\ \combadditive],
the only difference being that the subgradient of $F_{i_k}$ is taken at $z_k$ rather than $x_k$.

An important issue which affects the methods' effectiveness is the order in which the components $\{f_i,h_i\}$ are chosen for iteration. In this paper, we consider two possibilities:

\nitem{(1)} A {\it cyclic order\/}, whereby $\{f_i,h_i\}$ are taken up in the fixed deterministic order $1,\ldots,m$, so that $i_k$ is equal to ($k$ modulo $m$) plus 1. A contiguous block of iterations involving $\{f_1,h_1\},\ldots,\{f_m,h_m\}$ in this order and exactly once is called a {\it cycle\/}. We assume that the stepsize $\a_k$ is constant within a cycle (for all $k$ with $i_k=1$ we have $\a_k=\a_{k+1}\ldots=\a_{k+m-1}$). 

\nitem{(2)} A {\it randomized order based on uniform sampling\/}, whereby at each iteration a component pair $\{f_i,h_i\}$ is chosen randomly by sampling over all component pairs with a uniform distribution, independently of the past history of the algorithm.

\smskip
\pn It is essential to include all components in a cycle in the cyclic case, and to sample according to the uniform distribution in the randomized case, for otherwise some components will be sampled more often than others, leading to a bias in the convergence process. 

Another technique for
incremental methods, popular in neural network training practice, is to reshuffle randomly the order of the
component functions after each cycle. This alternative order selection scheme leads to convergence, like the preceding two. Moreover, this scheme has the nice property of allocating exactly one computation slot to each component in an $m$-slot cycle ($m$ incremental iterations). By comparison, choosing components by uniform sampling allocates one computation slot to each component {\it on the average\/}, but some components may not get a slot while others may get more than one. A nonzero variance in the number of slots that any fixed component gets within a cycle, may be detrimental to performance, and indicates that reshuffling randomly the order of the
component functions after each cycle works better; this is consistent with experimental observations shared with us by B.\ Recht (private communication). While it seems difficult to establish this fact analytically, a justification is suggested by the
view of the incremental method as a gradient-like method that uses as descent direction the true gradient at the start of the cycle plus an ``error" [due to the calculation of the component gradients at points intermediate within a cycle; cf.\ Eq.\ \graderr]. The error has apparently greater variance in the uniform sampling method than in the randomly shuffled order method (in fact the variance of the error would seem relatively larger as $m$ increases, although other factors such as variance of size of component gradients would also play a role). Heuristically, if the variance of the error is larger, the direction of descent deteriorates, suggesting slower convergence. In this paper, we will focus on the easier-to-analyze uniform sampling method, and show by analysis that it is superior to the cyclic order.

For the remainder of the paper, we denote by $F^*$ the optimal value of problem \combadditive:
$$F^*=\inf_{x\in X}F(x),$$
 and by $X^*$ the set of optimal solutions (which could be empty):
 $$X^*=\big\{x^*\mid x^*\in X,\,F(x^*)=F^*\big\}.$$
Also, for a nonempty closed convex set $X$, we denote by $\hbox{dist}(\cdot;X)$ the distance function given by
$$\hbox{dist}(x;X)=\min_{z\in X}\|x-z\|,\qquad x\in\rn.$$

In our convergence analysis of Section 4, we will use the following well-known 
theorem (see Neveu [Nev75], p.\ 33). We will use a much simpler deterministic version of the theorem in Section 3.

\xdef\supermartingale{\propn}\propnum\show{myproposition}

\texshopbox{
\proposition{\supermartingale: (Supermartingale
Convergence Theorem)} Let $Y_k$, $Z_k$, and $W_k$, $k=0,1,\ldots$, be three
sequences of random variables and let $\Fscr_k$, $k=0,1,\ldots$,
be sets of random variables such that $\Fscr_k\i \Fscr_{k+1}$
for all $k$. Suppose that:
\nitem{(1)}
The random variables $Y_k$, $Z_k$, and $W_k$ are nonnegative,
and are functions of the random variables in $\Fscr_k$.
\nitem{(2)} For each $k$, we have
$$E\bl\{ Y_{k+1}\mid \Fscr_k\br\}\le Y_k - Z_k + W_k.$$
\nitem{(3)} There holds, with probability 1, $\sum_{k=0}^\infty W_k < \infty$.
\smskip
\pn Then we have $\sum_{k=0}^\infty Z_k < \infty$, and the sequence
$Y_k$ converges to a nonnegative random variable $Y$, with probability 1.
}

\vskip-5mm
\section{Convergence for  Methods with Cyclic Order}
\vskip-2mm

\pn In this section, we discuss convergence under the cyclic order. We consider a randomized order in the next section. We focus on the sequence  $\{x_k\}$ rather than $\{z_k\}$, which need not lie within $X$ in the case of iterations \combincrsubgrat\ and \combincrsubgrbt\  when $X\ne\rn$. 
In summary, the idea is to show that the effect of taking subgradients of $f_i$ or $h_i$ at points near $x_k$ (e.g., at $z_k$ rather than at $x_k$) is inconsequential, and diminishes as the stepsize $\a_k$ becomes smaller, as long as some subgradients relevant to the algorithms are uniformly bounded in norm by some constant. This is similar to the convergence mechanism of incremental gradient methods described in Section 1.2. We use the following assumptions throughout the present section.

\xdef\assumptionthree{\assumptionn}\assumptionnum\show{myproposition}

\texshopbox{
\assumption{\assumptionthree: [For iterations \combincrsubgrt\ and  \combincrsubgrat]} 
There is a constant $c\in\re$ such that for all $k$
$$\max\big\{\|\suf_{i_k}(z_{k})\|,\|\suh_{i_k}(z_k)\|\big\}\le c.\xdef\ciscalara{\lab}\eqnum\show{target}$$
Furthermore, for all $k$ that mark the beginning of a cycle (i.e., all $k>0$ with $i_k=1$), we have
$$\max\big\{f_{j}(x_k)-f_{j}(z_{k+j-1}),\,h_{j}(x_k)-h_{j}(z_{k+j-1})\big\}\le c\,\|x_k-z_{k+j-1}\|,\qquad \forall\ j=1,\ldots,m.\xdef\ciscalarb{\lab}\eqnum\show{target}$$
}

\xdef\assumptionthreea{\assumptionn}\assumptionnum\show{myproposition}

\texshopbox{
\assumption{\assumptionthreea: [For iteration \combincrsubgrbt]} 
There is a constant $c\in\re$ such that for all $k$
$$\max\big\{\|\suf_{i_k}(x_{k+1})\|,\|\suh_{i_k}(x_k)\|\big\}\le c.\xdef\ciscalaraa{\lab}\eqnum\show{target}$$
Furthermore, for all $k$ that mark the beginning of a cycle (i.e., all $k>0$ with $i_k=1$), we have
$$\max\big\{f_{j}(x_k)-f_{j}(x_{k+j-1}),\,h_{j}(x_k)-h_{j}(x_{k+j-1})\big\}\le c\,\|x_k-x_{k+j-1}\|,\qquad \forall\ j=1,\ldots,m,\xdef\ciscalarbb{\lab}\eqnum\show{target}$$
$$f_{j}(x_{k+j-1})-f_{j}(x_{k+j})\le c\,\|x_{k+j-1}-x_{k+j}\|,\qquad \forall\ j=1,\ldots,m.\xdef\ciscalarcc{\lab}\eqnum\show{target}$$
}

Note that the condition \ciscalarb\ is satisfied if for each $i$ and $k$, there is a subgradient of $f_i$ at $x_k$ and a subgradient of $h_i$ at $x_k$, whose norms are bounded by $c$. Conditions that imply the preceding assumptions are:

\nitem{(a)} For algorithm \combincrsubgrt: $f_i$ and $h_i$ are Lipschitz continuous over the set $X$.

\nitem{(b)} For algorithms \combincrsubgrat\ and \combincrsubgrbt:  $f_i$ and $h_i$ are Lipschitz continuous over the entire space $\rn$. 

\nitem{(c)} For all algorithms \combincrsubgrt, \combincrsubgrat, and  \combincrsubgrbt: $f_i$ and $h_i$ are polyhedral [this is a special case of (a) and (b)].

\nitem{(d)} For all algorithms \combincrsubgrt, \combincrsubgrat, and  \combincrsubgrbt: The sequences $\{x_k\}$ and $\{z_k\}$ are bounded  [since then $f_i$ and $h_i$, being real-valued and convex, are Lipschitz continuous over any bounded set that contains $\{x_k\}$ and $\{z_k\}$].

\smskip

The following proposition provides a key estimate that reveals the convergence mechanism of our methods.\footnote{\dag}{\ninepoint  The original version of this report gave $\b={1\over m}+4$ for the case of algorithms \combincrsubgrt\ and \combincrsubgrat, and $\b={5\over m}+4$ for the case of algorithm \combincrsubgrbt, because a loose bound was used in the following calculation. The tighter version for  algorithm \combincrsubgrbt\ given here was prompted by an observation by M.\ Andersen and P.\ C.\ Hansen in Oct.\ 2013.} 

\xdef \lemmaprox{\propn}\propnum\show{myproposition}

\texshopbox{
\proposition{\lemmaprox:}
Let $\{x_k\}$
be the sequence generated by any one of the algorithms \combincrsubgrt-\combincrsubgrbt, with a cyclic order of component selection. Then for all $y\in X$ and all $k$ that mark the beginning of a cycle (i.e., all $k$ with $i_k=1$), we have
$$\|x_{k+m} - y\|^2 \le \|x_k - y\|^2
- 2\a_k\bl( F(x_{k}) - F(y)\br) + \a_k^2\b m^2c^2,
\xdef\target{\lab}\eqnum\show{target}$$
where $\b={1\over m}+4$.}
\proof
 We first prove the result for algorithms \combincrsubgrt\ and \combincrsubgrat, and then indicate the modifications necessary for algorithm \combincrsubgrbt.
 Using Prop.\ \propincrproxo(b), we have for all $y\in X$ and $k$,
$$\|z_k - y\|^2\le \|x_k - y\|^2
- 2\a_k \big(f_{i_k}(z_k)-f_{i_k}(y)\big).\xdef\longineqone{\lab}\eqnum\show{target}$$
Also, using the nonexpansion property of the
projection [i.e., $\big\|P_X(u)-P_X(v)\big\|\le \|u-v\|$ for all $u,v\in\rn$], 
the definition of subgradient, and Eq.\ \ciscalara,
we obtain for all $y\in X$ and $k$,
$$\eqalign{\|x_{k+1} - y\|^2
&= \bl\| P_X\big(z_k-\a_k \suh_{i_k}(z_k)\big)-y\br\|^2 \cr
&\le \|z_k-\a_k \suh_{i_k}(z_k)-y\|^2 \cr
&= \|z_k - y\|^2
- 2\a_k \suh_{i_k}(z_k)'(z_k - y) + \a_k^2\big\|\suh_{i_k}(z_k)\big\|^2 \cr
&\le \|z_k - y\|^2
- 2\a_k \big(h_{i_k}(z_k)-h_{i_k}(y)\big) + \a_k^2c^2.\cr}\xdef\longineqtwo{\lab}\eqnum\show{target}$$
Combining Eqs.\ \longineqone\ and \longineqtwo, and using the definition $F_j=f_j+h_j$, we have
$$\eqalign{\|x_{k+1}-y\|^2&\le \|x_k-y\|^2-2\a_k \bl(f_{i_k}(z_k)+h_{i_k}(z_k)-f_{i_k}(y)-h_{i_k}(y)\br)+ \a_k^2c^2\cr
&=\|x_k-y\|^2-2\a_k \bl(F_{i_k}(z_k)-F_{i_k}(y)\br)+ \a_k^2c^2.\cr}\xdef\firstest{\lab}\eqnum\show{target}$$

Let now $k$ mark the beginning of a cycle (i.e., $i_k=1$). Then at iteration $k+j-1$, $j=1,\ldots,m$, the selected components are $\{f_j,h_j\}$, in view of the assumed cyclic order.
We may thus replicate the preceding inequality with $k$ replaced by $k+1,\ldots,k+m-1$, and add to obtain
$$\|x_{k+m}-y\|^2\le \|x_k-y\|^2-2\a_k \sum_{j=1}^m\big(F_j(z_{k+j-1})-F_j(y)\br)+ m\a_k^2c^2,$$
or equivalently, since $F=\sum_{j=1}^mF_j$, 
$$\|x_{k+m}-y\|^2\le \|x_k-y\|^2-2\a_k \bl(F(x_k)-F(y)\br)+ m\a_k^2c^2+2\a_k\sum_{j=1}^m\big(F_j(x_k)-F_j(z_{k+j-1})\br).\xdef\ineqone{\lab}\eqnum\show{target}$$
The remainder of the proof deals with appropriately bounding the last term above.

 From Eq.\ \ciscalarb, we have for $j=1,\ldots,m$, 
$$F_j(x_k)-F_j(z_{k+j-1})\le 2c\,\|x_k-z_{k+j-1}\|.\xdef\ineqthreea{\lab}\eqnum\show{target}$$
We also have
$$\|x_k-z_{k+j-1}\|\le \|x_k-x_{k+1}\|+\cdots+\|x_{k+j-2}-x_{k+j-1}\|+\|x_{k+j-1}-z_{k+j-1}\|,\xdef\ineqthree{\lab}\eqnum\show{target}$$
and by the definition of the algorithms \combincrsubgrt\ and \combincrsubgrat, the nonexpansion property of the
projection,  and Eq.\ \ciscalara, each of the terms in the right-hand side above is bounded by $2\a_kc$, except for the last, which is bounded by $\a_kc$.  Thus Eq.\ \ineqthree\ yields $\|x_k-z_{k+j-1}\|\le \a_k(2j-1)c$, which together with Eq.\ \ineqthreea, shows that
$$F_j(x_k)-F_j(z_{k+j-1})\le 2\a_k c^2(2j-1).\xdef\ineqtwo{\lab}\eqnum\show{target}$$
Combining Eqs.\ \ineqone\ and \ineqtwo, we have 
$$\|x_{k+m}-y\|^2\le \|x_k-y\|^2-2\a_k \bl(F(x_k)-F(y)\br)+ m\a_k^2c^2+4\a_k^2c^2\sum_{j=1}^m(2j-1),$$
and finally
$$\|x_{k+m}-y\|^2\le \|x_k-y\|^2-2\a_k \bl(F(x_k)-F(y)\br)+ m\a_k^2c^2+4\a_k^2c^2m^2,$$
which is of the form \target\ with $\b={1\over m}+4$.

For the algorithm \combincrsubgrbt, a similar argument goes through using Assumption \assumptionthreea. In place of Eq.\ \longineqone, using the nonexpansion property of the
projection, 
the definition of subgradient, and Eq.\ \ciscalaraa,
we obtain for all $y\in X$ and $k\ge0$,
$$\|z_k - y\|^2\le \|x_k - y\|^2
- 2\a_k \big(h_{i_k}(x_k)-h_{i_k}(y)\big) + \a_k^2c^2,\eqnum\show{target}$$
while in place of Eq.\ \longineqtwo, using Prop.\ \propincrproxo(b), we have
$$\|x_{k+1} - y\|^2\le \|z_k - y\|^2
- 2\a_k \big(f_{i_k}(x_{k+1})-f_{i_k}(y)\big).\eqnum\show{target}$$
Combining these equations, in analogy with Eq.\ \firstest, we obtain
$$\eqalign{\|x_{k+1}-y\|^2&\le \|x_k-y\|^2-2\a_k \bl(f_{i_k}(x_{k+1})+h_{i_k}(x_k)-f_{i_k}(y)-h_{i_k}(y)\br)+ \a_k^2c^2\cr
&=\|x_k-y\|^2-2\a_k \bl(F_{i_k}(x_k)-F_{i_k}(y)\br)+ \a_k^2c^2+2\a_k\big(f_{i_k}(x_k)-f_{i_k}(x_{k+1})\big).\cr}\xdef\firstesta{\lab}\eqnum\show{target}$$

As earlier, we let $k$ mark the beginning of a cycle (i.e., $i_k=1$). We replicate the preceding inequality with $k$ replaced by $k+1,\ldots,k+m-1$, and add to obtain
[in analogy with Eq.\ \ineqone]
$$\eqalign{\|x_{k+m}-y\|^2&\le \|x_k-y\|^2-2\a_k \bl(F(x_k)-F(y)\br)+ m\a_k^2c^2\cr
&\ \ +2\a_k\sum_{j=1}^m\big(F_j(x_k)-F_j(x_{k+j-1})\br)+2\a_k\sum_{j=1}^m\big(f_j(x_{k+j-1})-f_j(x_{k+j})\big).\cr}\xdef\ineqonea{\lab}\eqnum\show{target}$$

We now bound the two sums in Eq.\ \ineqonea, using  Assumption \assumptionthreea. From Eq.\ \ciscalarbb, we have
$$F_j(x_k)-F_j(x_{k+j-1})\le 2c\|x_k-x_{k+j-1}\|\le 2c\big(\|x_k-x_{k+1}\|+\cdots+\|x_{k+j-2}-x_{k+j-1}\|\big),$$
and since by Eq.\ \ciscalaraa\ and the definition of the algorithm, each of the norm terms in the right-hand side above is bounded by $2\a_kc$,
$$F_j(x_k)-F_j(x_{k+j-1})\le 4\a_k c^2(j-1).$$
Also from Eqs.\  \ciscalaraa\ and \ciscalarcc, and the nonexpansion property of the projection, we have
$$f_j(x_{k+j-1})-f_j(x_{k+j})\le c\,\|x_{k+j-1}-x_{k+j}\|\le 2\a_kc^2.$$
Combining the preceding relations and adding, we obtain
$$\eqalign{2\a_k\sum_{j=1}^m\big(F_j(x_k)-F_j(x_{k+j-1})\br)&+2\a_k\sum_{j=1}^m\big(f_j(x_{k+j-1})-f_j(x_{k+j})\big)\le 8\a_k^2 c^2\sum_{j=1}^m(j-1)+4\a_k^2c^2m\cr
&=4\a_k^2 c^2(m^2-m)+4\a_k^2c^2m=\lf(4+{1\over m}\ri)\a_k^2c^2m^2,\cr}
$$
which together with Eq.\ \ineqonea, yields Eq.\ \target. \qed

Among other things, Prop.\ \lemmaprox\ guarantees that with a cyclic order, given the 
iterate
$x_k$ at the start of a cycle and
any point $y\in X$ having lower cost than $x_k$ (for example an optimal point), the algorithm yields a point
$x_{k+m}$ at the end of the cycle that will
be closer to $y$ than $x_k$, provided the stepsize $\a_k$ is less than 
$${2\bl(F(x_{k})-F(y)\br)\over \b m^2c^2}.$$
 In particular, for any $\e>0$ and assuming that there exists an optimal solution $x^*$, either we are within ${\a_k \b m^2c^2\over 2}+\e$ of the optimal value,
$$F(x_{k})\le F(x^*)+{\a_k \b m^2c^2\over 2}+\e,$$
or else the squared distance to $x^*$ will be strictly decreased by at least $2\a_k \e$,
$$\|x_{k+m}-x^*\|^2<\|x_k-x^*\|^2-2\a_k \e.$$
Thus, using Prop.\ \lemmaprox, we can provide various types of convergence results.  As an example, for a constant stepsize ($\a_k\equiv\a$), convergence can be established to a neighborhood of the optimum, which shrinks to 0 as $\a\to0$, as stated in the following proposition.

\xdef\proptwoone{\propn}\propnum\show{myproposition}

\texshopbox{
\proposition{\proptwoone:}
Let $\{x_k\}$
be the sequence generated by any one of the algorithms \combincrsubgrt-\combincrsubgrbt, with a cyclic order of component selection, and let the stepsize $\a_k$ be fixed
at some
positive constant $\a$.
\nitem{(a)} If $F^* = -\infty$, then
$$\liminf_{k\to\infty}F(x_{k}) = F^*.$$
\nitem{(b)} If $F^*> -\infty$, then
$$\liminf_{k\to\infty}F(x_{k})\le F^* + {\a \b m^2c^2\over2},$$
where\ $c$ and $\b$ are the constants of Prop.\ \lemmaprox.
}

\proof
We prove (a) and (b) simultaneously. 
If the result does not hold, there must exist an $\e>0$ such that
$$\liminf_{k\to\infty}F(x_{km}) - {\a \b m^2c^2\over2} - 2\e > F^*.$$
Let $\hat y\in X$ be such that
$$\liminf_{k\to\infty}F(x_{km})- {\a \b m^2c^2\over2} -
2\e \ge F(\hat y),$$  and let $k_0$ be large enough so that
for all $k\ge k_0$, we have
$$F(x_{km}) \ge \liminf_{k\to\infty}F(x_{km}) - \e.$$
By combining the preceding two relations, we obtain
for all $k\ge k_0$,
$$F(x_{km}) - F(\hat y)\ge {\a \b m^2c^2\over2} + \e. $$
Using Prop.\ \lemmaprox\ for the case where  $y=\hat y$ together with the above
relation, we
obtain for all $k\ge k_0$,
$$\|x_{(k+1)m} - \hat y\|^2 \le \|x_{km} - \hat y\|^2-2\a\big(F(x_{km}) - F(\hat y)\big)+ \b \a^2 m^2c^2\le
\|x_{km} - \hat y\|^2
- 2 \a\e.$$
This relation implies that for all $k\ge k_0$,
$$\|x_{(k+1)m} - \hat y\|^2
\le \|x_{(k-1)m} - \hat y\|^2
- 4 \a\e
\le \cdots
\le
\|x_{k_0} - \hat y\|^2 - 2(k+1-k_0) \a\e,$$
which cannot hold for $k$ sufficiently large -- a contradiction.
\qed

The next proposition gives an estimate of  the number of iterations needed
to guarantee a given level of optimality up to the threshold tolerance $\a\b m^2
c^2 /2$ of the preceding proposition.

\xdef\propdetrate{\propn}\propnum\show{myproposition}

\texshopbox{\proposition{\propdetrate:}Assume that $X^*$ is nonempty.
Let $\{x_k\}$ be  a sequence generated as in Prop.\ \proptwoone.
Then for $\e>0$, we have
$$\min_{0\le k\le N}F(x_{k}) \le F^* + {\a\b m^2 c^2 + \e\over 2},
\xdef\estnum{\lab}\eqnum\show{target}$$
where  $N$ is given by
$$
N=m\,\lf \lfloor {\hbox{dist}(x_0;X^*)^2 \over \a\e} \ri \rfloor.\xdef\nestnum{\lab}\eqnum\show{target}$$
}

\proof Assume,  to arrive at a contradiction,
that Eq.\ \estnum\ does not hold, so that
for all $k$ with $0\le km\le N$, we have
$$F(x_{km}) > F^* + {\a\b m^2 c^2 + \e\over 2}.$$
By using this relation in Prop.\ \lemmaprox\
with $\a_k$ replaced by $\a$ and $y$ equal to the vector of $X^*$ that is at minimum distance from $x_{km}$,
we obtain
for all $k$ with $0\le km\le N$,
$$\eqalign{\hbox{dist}(x_{(k+1)m};X^*)^2
&\le  \hbox{dist}(x_{km};X^*) ^2
   - 2\a\br( F(x_{km}) - F^* \bl) + \a^2\b m^2 c^2\cr
&\le \hbox{dist}(x_{km};X^*) ^2
   - (\a^2 \b m^2 c^2 + \a\e) + \a^2 \b m^2 c^2 \cr
&= \hbox{dist}(x_{km};X^*)^2 - \a\e.
\cr}$$
Adding  the above inequalities for $k=0,\ldots,{N\over m}$, we obtain
$$\hbox{dist}(x_{N+m};X^*) ^2 \le
 \hbox{dist}(x_0;X^*)^2 - \lf({N\over m}+1\ri)\a\e,
$$
so that
$$  \lf({N\over m}+1\ri)\a\e\le  \hbox{dist}(x_0;X^*)^2,$$
which contradicts the definition of $N$.
\qed

According to Prop.\ \propdetrate,  to achieve a cost function value within $O(\e)$ of the optimal, the term $\a\b m^2 c^2$ must also be of order $O(\e)$, so $\a$ must be of order $O(\e/m^2c^2)$, and from Eq.\ \nestnum, the number of necessary iterations $N$ is $O(m^3c^2/\e^2)$, and the number of necessary  cycles is $O\big((mc)^2/\e^2)\big)$. This is the same type of estimate as for the nonincremental subgradient method [i.e., $O(1/\e^2)$, counting a cycle as one iteration of the nonincremental method, and viewing $mc$ as a Lipschitz constant for the entire cost function $F$], and does not reveal any advantage for the incremental methods given here. However,  in the next section, we demonstrate a much more favorable iteration complexity estimate for the incremental methods that use a randomized order of component selection.

\subsubsection{Exact Convergence for a Diminishing Stepsize}

\pn We can also obtain an exact convergence result for the case where the stepsize $\a_k$ diminishes to zero. The idea is that with a constant stepsize $\a$ we can get to within an $O(\a)$-neighborhood of the optimum, as shown above, so with a diminishing stepsize $\a_k$, we should be able to reach an arbitrarily small neighborhood of the optimum. However, for this to happen, $\a_k$ should not be reduced too fast, and should 
satisfy $\sum_{k=0}^{\infty}\alpha_k = \infty$ (so that the method can
``travel" infinitely far
if necessary). 

\xdef\proptwotwo{\propn}\propnum\show{myproposition}

\texshopbox{
\proposition{\proptwotwo:}Let $\{x_k\}$
be the sequence generated by any one of the algorithms \combincrsubgrt-\combincrsubgrbt, with a cyclic order of component selection, and let the stepsize $\a_k$ satisfy
$$\lim_{k\to\infty}\alpha_k = 0,
\qquad \sum_{k=0}^{\infty}\alpha_k = \infty.$$
Then, 
$$\liminf_{k\to\infty} F(x_k) = F^{\,*}.$$
Furthermore, if $X^*$ is nonempty and 
$$\sum_{k=0}^{\infty}\alpha_k^2< \infty,$$
then $\{x_{k}\}$ converges to some $x^*\in X^*$.
}

\proof For the first part, it will be sufficient to show that $\liminf_{k\to\infty} F(x_{km}) = F^{\,*}.$ Assume, to arrive at a contradiction,
 that there exists
 an $\e>0$ such that 
 $$\liminf_{k\to\infty}F(x_{km}) - 2\e > F^{\,*}.$$
 Then there exists
 a point $\hat y\in X$ such that
 $$\liminf_{k\to\infty}F(x_{km}) - 2\e> F(\hat y).$$
 Let $k_0$ be large enough so that
 for all $k\ge k_0$, we have
 $$F(x_{km}) \ge \liminf_{k\to\infty}F(x_{km}) - \e.$$
 By combining the preceding two relations, we obtain
 for all $k\ge k_0$,
 $$F(x_{km}) - F(\hat y)> \e.$$
 By setting $y=\hat y$ in Prop.\ \lemmaprox,
 and by using the above relation,
 we have for all $k\ge k_0$,
 $$\|x_{(k+1)m} - \hat y\|^2
 \le \|x_{km} - \hat y\|^2 - 2\a_{km}\e +\b\a_{km}^2m^2 c^2
 = \|x_{km} - \hat y\|^2 - \a_{km}\lf( 2\e - \b\a_{km}m^2c^2\ri).$$
 Since $\a_k\to0$, without loss of generality, we may assume that $k_0$ is
 large enough so that
 $$2\e - \b\a_km^2 c^2\ge\e, \qquad \forall \ k\ge k_0.$$
 Therefore for all $k\ge k_0$, we have
 $$\|x_{(k+1)m} - \hat y\|^2
 \le \|x_{km} - \hat y\|^2 - \a_{km}\e
 \le \cdots
 \le \|x_{k_0m} - \hat y\|^2 - \e\sum_{\ell=k_0}^k \a_{\ell m},$$
 which cannot hold for $k$ sufficiently large. Hence $\liminf_{k\to\infty} F(x_{km}) = F^{\,*}.$
 
To prove the second part of the proposition, note that from Prop.\ \lemmaprox, for every $x^*\in X^*$ and $k\ge0$ we have
$$\|x_{(k+1)m} - x^*\|^2 \le \|x_{km} - x^*\|^2
- 2\a_{km}\bl( F(x_{km}) - F(x^*)\br) + \a_{km}^2\b m^2 c^2.\eqnum\show{dynrel}$$
The Supermartingale
Convergence Theorem (Prop.\ \supermartingale)\footnote{\dag}
{\ninepoint  Actually we use here a deterministic version/special case of the theorem, where $Y_k$, $Z_k$, and $W_k$ are nonnegative scalar sequences satisfying $Y_{k+1}\le Y_k-Z_k+W_k$ with $\sum_{k=0}^\infty W_k<\infty$. Then the sequence $Y_k$ must converge. This version is given with proof in many sources, including [BeT96] (Lemma 3.4), and [BeT00] (Lemma 1).}
 and the hypothesis $\sum_{k=0}^{\infty}\alpha_k^2 <\infty$, imply that $\big\{\|x_{km} - x^*\|\big\}$ converges for every $x^*\in X^*$. Since then $\{x_{km}\}$ is bounded, it has a limit point $\ol x\in X$ that satisfies 
 $$F(\ol x)=\liminf_{k\to\infty} F(x_{km}) = F^{\,*}.$$
This implies that $\ol x\in X^*$, so it follows that $\big\{\|x_{km} - \ol x\|\big\}$ converges, and that the entire sequence $\{x_{km}\}$ converges to $\ol x$ (since $\ol x$ is a limit point of $\{x_{km}\}$). 
  
Finally, to show that the entire sequence  $\{x_{k}\}$ also converges to $\ol x$, note that from Eqs.\ \ciscalara\ and \ciscalaraa, and the form of the iterations \combincrsubgrt-\combincrsubgrbt, we have $\|x_{k+1}-x_k\|\le 2\a_k c\to0$. Since $\{x_{km}\}$ converges to $\ol x$, it follows that $\{x_{k}\}$ also converges to $\ol x$. \qed

\vskip-8mm
\section{Convergence for  Methods with Randomized Order}

\vskip-2mm
\xdef\propthreeone{\propn}\propnum\show{myproposition}

\pn 
In this section, we discuss convergence for the randomized component selection order and a constant stepsize $\a$. The randomized versions of iterations \combincrsubgrt, \combincrsubgrat, and \combincrsubgrbt,  are
$$z_k=P_X\big(x_k-\a \suf_{\o_k}(z_k)\big),\qquad x_{k+1}=P_X\big(z_k-\a \suh_{\o_k}(z_k)\big),\xdef\random{\lab}\eqnum\show{twoo}$$
$$z_k=x_k-\a \suf_{\o_k}(z_k),\qquad x_{k+1}=P_X\big(z_k-\a \suh_{\o_k}(z_k)\big),\xdef\randoma{\lab}\eqnum\show{twoo}$$
$$z_k=P_X\big(x_k-\a \suh_{\o_k}(z_k)\big),\qquad x_{k+1}=z_k-\a \suf_{\o_k}(x_{k+1}),\xdef\randomb{\lab}\eqnum\show{twoo}$$
respectively, where  $\{\o_k\}$ is a sequence of random variables,
taking values from the index set $\{1,\ldots,m\}$.

We assume the following throughout the present section.

\xdef\assumptionthreeone{\assumptionn}\assumptionnum\show{myproposition}

\texshopbox{
\assumption{\assumptionthreeone: [For iterations \random\ and \randoma]} 
\nitem{(a)} $\{\o_k\}$ is a sequence of
random
variables, each
uniformly distributed over $\{1,\ldots,m\}$, and such that for each $k$, $\o_k$ is
independent of the past history $\{x_k,z_{k-1},x_{k-1},\ldots,z_0,x_0\}$.
\nitem{(b)} There is a constant $c\in\re$
such that for all $k$, we have with probability 1
$$\max\big\{\|\suf_{i}(z_k^i)\|,\,\|\suh_{i}(z_k^i)\|\big\}\le c,\qquad \forall\ i=1,\ldots,m,\xdef\ciscalarc{\lab}\eqnum\show{target}
$$
$$\max\big\{f_{i}(x_k)-f_{i}(z_k^i),\,h_{i}(x_k)-h_{i}(z_k^i)\big\}\le c\|x_k-z_k^i\|,\qquad \forall\ i=1,\ldots,m,\xdef\ciscalard{\lab}\eqnum\show{target}$$
where $z_k^i$ is the result of the proximal iteration, starting at $x_k$ if $\o_k$ would be $i$, i.e.,$$z_k^i=\arg\min_{x\in X}\lf\{f_i(x)+{1\over 2\a_k}\|x-x_k\|^2\ri\},\old{\eqnum\show{target}}$$
in the case of iteration \random, and
$$z_k^i=\arg\min_{x\in \rn}\lf\{f_i(x)+{1\over 2\a_k}\|x-x_k\|^2\ri\},\old{\eqnum\show{target}}$$
in the case of iteration \randoma.}

\xdef\assumptionthreeonea{\assumptionn}\assumptionnum\show{myproposition}

\texshopboxnb{
\assumption{\assumptionthreeonea: [For iteration \randomb]} 
\nitem{(a)} $\{\o_k\}$ is a sequence of
random
variables, each
uniformly distributed over $\{1,\ldots,m\}$, and such that for each $k$, $\o_k$ is
independent of the past history $\{x_k,z_{k-1},x_{k-1},\ldots,z_0,x_0\}$.
\nitem{(b)} There is a constant $c\in\re$
such that for all $k$, we have with probability 1
$$\max\big\{\|\suf_{i}(x_{k+1}^i)\|,\,\|\suh_{i}(x_k)\|\big\}\le c,\qquad \forall\ i=1,\ldots,m,\xdef\ciscalarcc{\lab}\eqnum\show{target}
$$
$$f_{i}(x_k)-f_{i}(x_{k+1}^i)\le c\|x_k-x_{k+1}^i\|,\qquad \forall\ i=1,\ldots,m,\xdef\ciscalardd{\lab}\eqnum\show{target}$$}\texshopboxnt{\item{}
where $x_{k+1}^i$ is the result of the iteration, starting at $x_k$ if $\o_k$ would be $i$, i.e.,
$$x_{k+1}^i=P_X\big(z_k^i-\a_k \suf_i(x_{k+1}^i)\big),\old{\eqnum\show{target}}$$
with
$$z_k^i=x_k-\a_k \suh_i(x_k).\old{\eqnum\show{target}}$$
}

Note that condition \ciscalard\ is satisfied if there exist subgradients of $f_i$ and $h_i$ at $x_k$ with norms less or equal to $c$. Thus the conditions \ciscalarc\ and \ciscalard\ are similar, the main difference being that the first applies to ``slopes" of $f_i$ and $h_i$ at $z_k^i$ while the second applies to the ``slopes" of $f_i$ and $h_i$ at $x_k$.   
As in the case of Assumption \assumptionthree, these conditions are guaranteed by Lipschitz continuity assumptions on $f_i$ and $h_i$. The convergence analysis of the randomized algorithms of this section is somewhat more complicated than the one of the cyclic order counterparts, and relies on the Supermartingale Convergence Theorem. The following proposition deals with the case of a constant stepsize, and parallels Prop.\ \proptwoone\ for the cyclic order case.

\texshopbox{
\proposition{\propthreeone:}
Let $\{x_k\}$
be the sequence generated by one of the randomized incremental methods \random-\randomb, and let the stepsize $\a_k$ be fixed at some positive constant $\a$.
\nitem{(a)} If $F^*=-\infty$, then with probability 1
$$\inf_{k\ge0} F(x_k) = F^*.$$
\nitem{(b)} If $F^*>-\infty$, then with probability 1
$$\inf_{k\ge0} F(x_k) \le F^* + {\a \b mc^2\over2},$$
where $\b=5$.
}

\proof  Consider first algorithms  \random\ and \randoma. By adapting the proof argument of Prop.\ \lemmaprox\ with $F_{i_k}$ replaced by
$F_{\o_k}$ [cf.\ Eq.\ \firstest], we have
$$\|x_{k+1} - y\|^2 \le \|x_k - y\|^2
-2\a \bl(F_{\o_k}(z_k)-F_{\o_k}(y)\br)+ \a^2 c^2,\quad  \forall\ y\in
X,\quad
k\ge0.$$  
By taking the conditional expectation
with respect to ${\cal F}_k = \{x_k,z_{k-1},\ldots,z_0,x_0\}$, and using the fact that $\o_k$ takes the values
$i=1,\ldots,m$ with equal probability $1/m$, we obtain  for all $y\in X$ and $k$,
$$\eqalign{E\bl\{ \|x_{k+1} - y\|^2\mid {\cal F}_k \br\}
&\le
\|x_k - y\|^2 - 2\a E\bl\{ F_{\o_k}(z_k) -F_{\o_k}(y)\mid {\cal F}_k \br\} +
\a^2c^2\cr
&= \|x_k - y\|^2 - {2\a\over m}\sum_{i=1}^m\bl( F_i(z_k^i) -F_i(y) \br) +
\a^2c^2\cr
&= \|x_k - y\|^2 - {2\a\over m} \bl( F(x_k) -F(y) \br)+ {2\a\over m}\sum_{i=1}^m\bl(F_i(x_k)-F_i(z_k^i) \br)+
\a^2c^2.\cr}
\xdef\basicest{\lab}\eqnum\show{new}$$
By 
using Eqs.\ \ciscalarc\ and \ciscalard,
$$\sum_{i=1}^m\bl(F_i(x_k)-F_i(z_k^i) \br)\le 2c\sum_{i=1}^m\|x_k-z_k^i\|=2c\a\sum_{i=1}^m\|\suf_i(z_k^i)\|\le 2m\a c^2.$$
By combining the preceding two relations, we obtain
$$\eqalign{E\bl\{ \|x_{k+1} - y\|^2\mid {\cal F}_k \br\}&
\le
\|x_k - y\|^2- {2\a\over m}\bl( F(x_k) -F(y) \br)+4 \a^2c^2+\a^2c^2\cr
&=
\|x_k - y\|^2- {2\a\over m}\bl( F(x_k) -F(y) \br)+\b \a^2c^2,\cr}\xdef\new{\lab}\eqnum\show{kapax}$$
where $\b= 5$.

The preceding equation holds also for algorithm \randomb. To see this note that Eq.\ \firstesta\ yields for all $y\in X$
$$\|x_{k+1}-y\|^2\le \|x_k-y\|^2-2\a \bl(F_{\o_k}(x_k)-F_{\o_k}(y)\br)+ \a^2c^2+2\a\big(f_{\o_k}(x_k)-f_{\o_k}(x_{k+1})\big),\eqnum\show{target}$$
and similar to Eq.\ \basicest, we obtain
$$E\bl\{ \|x_{k+1} - y\|^2\mid {\cal F}_k \br\}
\le
\|x_k - y\|^2 - {2\a\over m} \bl( F(x_k) -F(y) \br)+ {2\a\over m}\sum_{i=1}^m\bl(f_i(x_k)-f_i(x_{k+1}^i) \br)+
\a^2c^2.
\eqnum\show{new}$$
From Eq.\ \ciscalardd, we have
$$f_i(x_k)-f_i(x_{k+1}^i)\le c\|x_k-x_{k+1}^i\|,$$
and from Eq.\ \ciscalarcc\ and the nonexpansion property of the projection, $$\|x_k-x_{k+1}^i\|\le \big\|x_k-z_k^i+\a \suf_i(x_{k+1}^i)\big\| = \big\|x_k-x_k+\a \suh_i(x_k)+\a \suf_i(x_{k+1}^i)\big\|\le 2\a c.$$
Combining the preceding inequalities, we obtain Eq.\ \new\
with $\b= 5$.

Let us fix a positive scalar $\g$, consider the level set $L_\g$ defined by
$$L_\g =\cases{\lf\{ x\in X \mid F(x) < -\g+1 + {\a\b mc^2\over2} \ri\}&if
$F^*=-\infty$,\cr
\lf\{ x\in X \mid F(x) < F^*  + {2\over \g}+ {\a\b mc^2\over2}\ri\}&if
$F^*>-\infty$,\cr}$$
and let $y_\g\in X$ be such that
$$F(y_\g)=\cases{ -\g&if $F^*=-\infty$,\cr
F^*+{1\over \g}&if $F^*>-\infty$.\cr}$$
Note that $y_\g\in L_\g$ by construction. 
Define a new process $\{\hat
x_k\}$ that is identical to
$\{x_k\}$, except that
once $x_k$ enters the level set $L_\g$, the process terminates with $\hat
x_k = y_\g$.
We will now argue that for any fixed $\g$, $\{\hat x_k\}$ (and hence also $\{x_k\}$) will
eventually
enter $L_\g$, which will prove both parts (a) and (b).

Using Eq.\ \new\ with $y=y_\g$, we have
$$E\bl\{ \|\hat x_{k+1} - y_\g\|^2\mid {\cal F}_k \br\}
\le \|\hat x_k - y_\g\|^2 - {2\a\over m}\bl( F(\hat x_k) - F(y_\g) \br) +
\b\a^2c^2,$$
from which
$$E\bl\{ \|\hat x_{k+1} - y_\g\|^2\mid {\cal F}_k \br\}
\le \|\hat x_k - y_\g\|^2 - v_k,
\xdef\martineq{\lab}\eqnum\show{kapax}$$
where
$$v_k = \cases{ {2\a\over m}\bl(F(\hat x_k) - F(y_\g) \br) -\b\a^2c^2
& if $\hat x_k\notin L_\g$, \cr
0 & if $\hat x_k = y_\g$.\cr}$$
The idea of the subsequent argument is to show that as long as
$\hat x_k\notin
L_\g$, the scalar $v_k$ (which is a measure of progress) is strictly
positive and
bounded away from 0.

\smskip
\pn (a) Let $F^*=-\infty$. Then if $\hat x_k \notin L_\g$, we have
$$\eqalign{v_k &= {2\a\over m} \bl(F(\hat x_k) - F(y_\g) \br) -\b\a^2 c^2\cr
&\ge {2\a\over m} \lf(-\g+1 + {\a\b mc^2\over2} + \g \ri) - \b\a^2 c^2\cr
&={2\a\over m}.\cr}$$
Since $v_k=0$ for $\hat x_k \in L_\g$, we have $v_k\ge0$ for all $k$, and by
Eq.\ \martineq\ and the Supermartingale Convergence Theorem (cf.\ Prop.\ \supermartingale),
$\sum_{k=0}^\infty v_k<\infty$
implying that $\hat x_k \in L_\g$ for
sufficiently large $k$,  with probability 1. Therefore,
in the original process we have
$$\inf_{k\ge0} F(x_k) \le -\g + 1 + {\a \b mc^2\over2}$$
with probability 1.   Letting $\g\to\infty$,
we obtain $\inf_{k\ge0} F(x_k)=-\infty$ with probability 1.

\smskip
\pn (b) Let $F^*>-\infty$. Then if $\hat x_k \notin L_\g$, we have
$$\eqalign{v_k &= {2\a\over m} \bl(F(\hat x_k) - F(y_\g) \br) - \b\a^2 c^2\cr
&\ge {2\a\over m} \lf(F^* +{2\over \g}  + {\a \b mc^2\over2}
- F^* - {1\over \g}\ri) - \b \a^2 c^2\cr
&= {2\a \over m\g}.\cr}$$
Hence, $v_k\ge0$ for all $k$, and by
the Supermartingale Convergence Theorem, we have
$\sum_{k=0}^\infty v_k<\infty$
implying that $\hat x_k \in L_\g$ for
sufficiently large $k$, so that
in the original process,
$$\inf_{k\ge0} F(x_k) \le F^* +{2\over \g}+ {\a \b mc^2\over2}$$
with probability 1. Letting $\g\to\infty$,
we obtain $\inf_{k\ge0} F(x_k)\le F^*+ {\a \b mc^2/2}$. \qed

By comparing  Prop.\ \propthreeone(b) with Prop.\
\proptwoone(b), we see that when $F^*>-\infty$ and the stepsize $\a$ is constant,
the randomized methods \random, \randoma, and \randomb,  have a better error
bound (by
a factor $m$) than their nonrandomized
counterparts. In fact an example given in p.\ 514 of [BNO03] for the incremental subgradient method can be adapted to show that the bound of Prop.\ \proptwoone(b) is tight in the sense that for a bad problem/cyclic order we have $\lim\inf_{k\to\infty} F(x_k)-F^* = O(\a m^2c^2)$. By contrast the randomized method will get to within $O(\a m c^2)$ with probability 1 for any problem, according to Prop.\ \propthreeone(b). Thus with the randomized algorithm we do not run the risk of choosing by accident a bad cyclic order. A related result is provided by the following proposition, which
should be compared with Prop.\ \propdetrate\ for the nonrandomized methods.

\xdef\proprndrate{\propn}\propnum\show{myproposition}

\texshopbox{
\proposition{\proprndrate:}
Assume that $X^*$ is nonempty. Let $\{x_k\}$
be a sequence generated as in Prop.\ \propthreeone. Then for any positive scalar $\e$,
we have with probability 1
$$\min_{0\le k\le N} F(x_k) \le F^* + {\a \b m c^2
+\e\over2},\eqnum\show{first}$$where $N$ is a random variable with
$$E\bl\{N\br\} \le
m\,{\hbox{dist}(x_0;X^*)^2 \over \a\e}.\eqnum\show{first}
$$
}

\proof Let $\hat y$ be some fixed vector in  $X^*$. Define a new process $\{\hat x_k\}$ which is identical to $\{x_k\}$ except that
once $x_k$ enters the level set 
$$L= \lf\{x\in X \ \Big|\ F(x) < F^* + {\a \b m c^2+ \e\over 2}\ri\},
$$
the process  $\{\hat x_k\}$ terminates at $\hat y$. Similar to the
proof of Prop.\ \propthreeone\ [cf.\ Eq.\ \new\ with $y$ being the closest point of $\hat x_k$ in $X^*$], for the
process
$\{\hat x_k\}$  we obtain  for all $k$,
$$\eqalign{E\bl\{  \hbox{dist}(\hat x_{k+1};X^*)^2 \mid  \Fscr_k\br\} &\le E\bl\{  \|\hat x_{k+1}-y\|^2 \mid  \Fscr_k\br\}\cr
& \le
\hbox{dist}(\hat x_k;X^*)^2 - {2\a\over m} \bl( F(\hat x_k) - F^* \br) + \b\a^2 c^2\cr
&= \hbox{dist}(\hat x_k;X^*)^2 - v_k,\cr}
\xdef\fa{\lab}\eqnum\show{first}
$$
where $\Fscr_k  = \{x_k,z_{k-1},\ldots,z_0,x_0\}$ and
$$v_k=\cases{
{2\a\over m}\bl( F(\hat x_k) - F^* \br) - \b\a^2 c^2
& if $\hat x_k\not\in L$, \cr
0 & otherwise.\cr}$$
In the case where $\hat x_k\not\in L$, we have
$$v_k
\ge
{2\a\over m} \lf(F^* + {\a \b mc^2 +\e\over2} - F^* \ri) -\b \a^2 c^2\
= {\a\e \over m}.
\xdef\fe{\lab}\eqnum\show{first}$$
By the Supermartingale Convergence Theorem (cf.\ Prop.\ \supermartingale), from Eq.\ \fa\ we have
$$
\sum_{k=0}^\infty v_k <\infty
$$
with probability 1,
so that $v_k =0$ for all $k\ge N$, where $N$ is a random variable.
Hence $\hat x_N\in L$ with probability 1,
implying that in the original process we have
$$\min_{0\le k\le N} F(x_k) \le F^* + {\a \b mc^2 +\e\over 2}
$$
with probability 1. Furthermore, by taking the total expectation in Eq.\ \fa,
we obtain for all $k$,
$$E \bl\{ \hbox{dist}(\hat x_{k+1};X^*)^2 \br\}
\le
E \bl\{ \hbox{dist}(\hat x_k;X^*)^2 \br\} - E\{ v_k \}
\le
\hbox{dist}(\hat x_0;X^*)^2 - E\lf\{\sum_{j=0}^k v_j \ri\},$$
where in the last inequality we use the facts $\hat x_0= x_0$ and
$
E \bl\{\hbox{dist}(\hat x_0;X^*)^2 \br\}=\hbox{dist}(\hat x_0;X^*)^2.
$
Therefore, letting $k\to\infty$, and using the definition of $v_k$ and Eq.\ \fe,
$$ \hbox{dist}(\hat x_0;X^*)^2 \ge E\lf\{\sum_{k=0}^\infty v_k \ri\}
= E\lf\{\sum_{k=0}^{N-1} v_k \ri\} \ge E\lf\{ {N \a\e\over m}\ri\}
=  {\a\e\over m} E\bl\{ N \br\}.
$$
{\bf Q.E.D.}
\smskip

Like Prop.\ \propthreeone, a comparison of Props.\ \propdetrate\ and \proprndrate\ again suggests an advantage for the randomized methods: compared to their deterministic counterparts, they achieve a much smaller error tolerance (a factor of $m$),  in the same {\it expected} number of iterations. Note, however, that the preceding assessment is based on upper bound estimates, which may not be sharp on a given problem [although the bound of Prop.\ \proptwoone(b) is tight with a worst-case problem selection as mentioned earlier; see [BNO03], p.\ 514].  Moreover, the comparison based on worst-case values versus expected values may not be strictly valid. In particular, while Prop.\ \propdetrate\ provides an upper bound estimate on $N$, Prop.\ \proprndrate\ provides an upper bound estimate on $E\{N\}$, which is not quite the same.

Finally for the case of a diminishing stepsize, let us give the following proposition, which parallels Prop.\ \proptwotwo\ for the cyclic order.

\xdef\propthreetwo{\propn}\propnum\show{myproposition}

\texshopbox{
\proposition{\propthreetwo:}
Let $\{x_k\}$
be the sequence generated by one of the randomized incremental methods \random-\randomb, and let  the stepsize $\a_k$ satisfy
$$\lim_{k\to\infty}\alpha_k = 0,
\qquad \sum_{k=0}^{\infty}\alpha_k = \infty.$$
Then, with probability 1,
$$\liminf_{k\to\infty} F(x_k) = F^{\,*}.$$
Furthermore, if $X^*$ is nonempty and 
$$\sum_{k=0}^{\infty}\alpha_k^2< \infty,$$
then $\{x_k\}$ converges to some $x^*\in X^*$ with probability 1.
}

\proof The proof of the first part is nearly identical to the corresponding part of Prop.\ \proptwotwo. To prove the second part, similar to the proof of Prop.\ \propthreeone, we obtain
for all $k$ and all $x^*\in X^*$,
$$E\bl\{ \|x_{k+1} - x^*\|^2\mid {\cal F}_k \br\}
\le
\|x_k - x^*\|^2 - {2\a_k\over m}\bl( F(x_k) - F^* \br) +
\b\a_k^2c^2\eqnum\show{lifk}
$$
[cf.\  Eq.\ \new\ with $\a$ and $y$ replaced with $\a_k$ and $x^*$,
respectively],
where ${\cal F}_k =  \{x_k,z_{k-1},\ldots,z_0,x_0\}$.
By the Supermartingale Convergence Theorem (Prop.\ \supermartingale), for
each $x^*\in X^*$, there is a set $\Omega_{x^*}$ of sample paths of probability 1 such that for 
each sample path in $\Omega_{x^*}$
$$\sum_{k=0}^\infty {2\a_k\over m}\bl( F(x_k) - F^* \br)<\infty,
\xdef\lifk{\lab}\eqnum\show{lifk}$$
and the sequence $\{\|x_k-x^*\|\}$ converges.

\old{for
each $x^*\in
X^*$, we have  for all sample paths in a set $\Omega_{x^*}$ of probability 1
$$\sum_{k=0}^\infty {2\a_k\over m}\bl( F(x_k) - F^* \br)<\infty,
\xdef\lifk{\lab}\eqnum\show{lifk}$$
and the sequence $\{\|x_k-x^*\|\}$ converges.}

Let $\{v_i\}$ be a countable subset of the relative interior
$\hbox{ri}(X^*)$ that is dense in $X^*$
[such a set exists since $\hbox{ri}(X^*)$ is a relatively open subset of
the affine hull of
$X^*$; an example of such a set is the intersection of $X^*$ with the set
of vectors of the form
$x^*+\sum_{i=1}^pr_i\xi_i$, where
$\xi_1,\ldots,\xi_p$ are basis vectors for the affine hull of $X^*$ and
$r_i$ are rational numbers]. Let also $\Omega_{v_i}$ be the set of sample paths defined earlier that corresponds to $v_i$.
The intersection
$$\ol \Omega=\cap_{i=1}^\infty \Omega_{v_i}$$
has probability 1, since its complement $\ol \Omega^c$ is equal to
$\cup_{i=1}^\infty  \Omega_{v_i}^c$ and
$$\hbox{Prob}\lf(\cup_{i=1}^\infty 
\Omega_{v_i}^c\ri)\le \sum_{i=1}^\infty\hbox{Prob}\lf(
\Omega_{v_i}^c\ri)=0.$$

For each sample path in $\ol \Omega$, all the sequences $\{\|x_k-v_i\|\}$ converge so
that $\{x_k\}$ is bounded, while by the first part of the proposition [or Eq.\ \lifk] $\liminf_{k\to\infty}F(x_k)=F^*$. Therefore, $\{x_k\}$ has a limit point $\ol x$ in $X^*$. Since $\{v_i\}$ is dense in $X^*$, for every $\e>0$ there exists $v_{i(\e)}$ such that $\|\ol x-v_{i(\e)}\|<\e$. Since the sequence $\{\|x_k-v_{i(\e)}\|\}$ converges and $\ol x$ is a limit point of $\{x_k\}$, we have $\lim_{k\to\infty}\|x_k-v_{i(\e)}\|<\e$, so that 
$$\limsup_{k\to\infty}\|x_k-\ol x\|\le \lim_{k\to\infty}\|x_k-v_{i(\e)}\|+\|v_{i(\e)}-\ol x\|<2\e.$$
By taking $\e\to0$, it follows that $x_k\to\ol x$. \qed

\vskip  -4mm
\section{Some Applications}
\vskip  -2mm

\pn In this section we illustrate our methods in the context of two types of practical applications, and discuss relations with known algorithms.

\subsection{Regularized Least Squares}

\pn 
Let us consider least squares problems,  involving minimization of a sum of quadratic component functions $f_i(x)$ that correspond to errors between data and the output of a model that is parameterized by a vector $x$. Often a convex regularization function $R(x)$ is added to the least squares objective, to induce desirable properties of the solution. This gives rise to problems of the form
$$\eqalign{\hbox{\rm minimize}\quad &
R(x)+{1\over 2}\sum_{i=1}^m(c_i'x-d_i)^2\cr
\hbox{\rm subject to\ \ }
&x\in \rn,\cr}\xdef\regls{\lab}\eqnum\show{first}$$
where $c_i$ and $d_i$ are given vectors and scalars, respectively, and $\g$ is a positive scalar. 
When $R$ is differentiable (e.g., quadratic), and either $m$ is very large or the data $(c_i,d_i)$ become available sequentially over time, it makes sense to consider incremental gradient methods, which have a long history of applications over the last 50 years, starting with the 
Widrow-Hoff least mean squares (LMS) method [WiH60].

The classical type of regularization involves a quadratic function $R$ (as in classical regression and the LMS method), but nondifferentiable regularization functions have become increasingly important recently. On the other hand, to apply our incremental methods, a quadratic $R$ is not essential. What is important is that $R$ has a simple form that facilitates the use of proximal algorithms, such as for example a separable form, so that the proximal iteration on $R$ is simplified through decomposition.  As an example, consider the {\it $\ell_1$-regularization problem\/}, where
$$R(x)=\g \|x\|_1=\g \sum_{j=1}^n|x^j|,\xdef\lonepen{\lab}\eqnum\show{first}$$
$\g$ is a positive scalar and $x^j$ is the $j$th coordinate of $x$. Then the proximal iteration
$$z_k=\arg\min_{x\in\rn}\lf\{\g\,\|x\|_1+{1\over 2\a_k}\|x-x_k\|^2\ri\}$$
decomposes into the $n$ one-dimensional minimizations
$$z_k^j=\arg\min_{x^j\in\re}\lf\{\g \,|x^j|+{1\over 2\a_k}|x^j-x^j_k|^2\ri\},\qquad j=1,\ldots,n,$$
and can be done in closed form
$$z_k^j=\cases{x_k^j-\g \a_k&if $ \g\a_k\le x_k^j$,\cr
0&if $-\g\a_k<x_k^j< \g\a_k$,\cr
x_k^j+\g \a_k&if $ x_k^j\le -\g\a_k$,\cr}\qquad j=1,\ldots,n.\xdef\proxlonepen{\lab}\eqnum\show{first}$$
We refer to Figueiredo, Nowak, and Wright [FNW07], Wright, Nowak, and Figueiredo [WNF08], Beck and Teboulle [BeT10], and the references given there, for a discussion of a broad variety of applications in estimation and signal processing problems, where nondifferentiable regularization functions play an important role. 

We now note that the incremental algorithms of this paper are well-suited for solution of $\ell_1$-regularization problems of the form \regls-\lonepen. For example, the $k$th incremental iteration may consist of selecting a data pair $(c_{i_k},d_{i_k})$ and performing a proximal iteration of the form \proxlonepen\ to obtain $z_k$, followed by a gradient iteration on the component $\half (c_{i_k}'x-d_{i_k})^2$, starting at $z_k$:
$$x_{k+1}=z_k-\a_kc_{i_k}(c_{i_k}'z_k-d_{i_k}).$$
This algorithm is the special case of the algorithms \combincrsubgrt-\combincrsubgrbt\ (here $X=\rn$, and all three algorithms coincide), with $f_i(x)$ being $\g\|x\|_1$ (we use $m$ copies of this function) and $h_i(x)=\half(c_i'x-d_i)^2$. It can be viewed as an incremental version of a popular class of algorithms in signal processing, known as iterative shrinkage/thresholding  (see Chambolle et.\ al. [CDL98], Figueiredo and Nowak [FiN03], Daubechies, Defrise, and Mol [DDM04], Combettes and Wajs [CoW05],  Bioucas-Dias and Figueiredo [BiF07], Elad, Matalon, and Zibulevsky [EMZ07], Beck and Teboulle [BeT09], [BeT10]). 
Our methods bear the same relation to this class of algorithms as the LMS method bears to gradient algorithms for the classical linear least squares problem with quadratic regularization function.

Finally, let us note that as an alternative, the proximal iteration  \proxlonepen\ could be replaced by a proximal iteration on $\g\,|x^j|$ for some selected index $j$, with all indexes selected cyclically in incremental iterations. Randomized selection of  the data pair $(c_{i_k},d_{i_k})$ would also be interesting, particularly in contexts where the data has a natural stochastic interpretation.

\subsection{Iterated Projection Algorithms}

\pn A feasibility problem that arises in many contexts involves finding a point with certain properties within a set intersection $\cap_{i=1}^mX_i$, where each $X_i$ is a closed convex set. For the case where $m$ is large and each of the sets $X_i$ has a simple form, incremental methods that make successive projections on the component sets $X_i$ have a long history (see e.g., Gubin, Polyak, and Raik [GPR67], and recent papers such as Bauschke [Bau01], Bauschke, Combettes, and Kruk [BCL06], and Cegielski and Suchocka [CeS08], and  their  bibliographies). We may  consider the following generalized version of the classical feasibility problem,
$$\eqalign{\hbox{\rm minimize}\quad &
f(x)\cr
\hbox{\rm subject to\ \ }
&x\in \cap_{i=1}^mX_i,\cr}\xdef\iterprojprob{\lab}\eqnum\show{first}$$
where $f:\rn\mapsto\re$ is  a convex cost function, and the method
$$x_{k+1}=P_{X_{i_k}}\big(x_k-\a_k \tl \gr f(x_k)\big),\eqnum\show{first}$$
where the index $i_k$ is chosen from $\{1,\ldots,m\}$ according to a randomized rule.  Incremental algorithms for problem \iterprojprob, which bear some relation with ours have been recently proposed by Nedi\'c [Ned10]. 
Actually, the algorithm of [Ned10] involves an additional projection on a special set $X_0$ at each iteration, but for simplicity we will take $X_0=\rn$. The incremental approach is particularly well-suited for problems of the form \iterprojprob\ where the sets $X_i$ are not known in advance, but are revealed as the algorithm progresses.

While the problem \iterprojprob\ does not involve a sum of component functions, it may be converted into one that does by using an exact penalty function. In particular, consider the problem
$$\eqalign{\hbox{\rm minimize}\quad &
f(x)+\g  \sum_{i=1}^m\hbox{dist}(x;X_i)\cr
\hbox{\rm subject to\ \ }
&x\in \rn,\cr}\xdef\iterprojpen{\lab}\eqnum\show{first}$$
where $\g $ is a positive penalty parameter. Then for $f$ Lipschitz continuous and $\g$ sufficiently large, problems \iterprojprob\ and \iterprojpen\ are equivalent. We show this for the case where $m=1$ and then we generalize.

\xdef\propgenexact{\propn}\propnum\show{myproposition}

\texshopbox{\proposition{\propgenexact:} 
Let $f:Y\mapsto \re$ be a function defined on a subset $Y$ of $\rn$, and let $X$ be a nonempty closed subset of $Y$.
Assume that $f$ is Lipschitz continuous over $Y$ with constant $L$, i.e.,
$$\bl|f(x)-f(y)\br|\le L\|x-y\|,\qquad \forall\ x,y\in Y,$$
and let $\g $ be a scalar with
$\g > L$. Then the set of minima of $f$ over
$X$ coincides with the set of minima of 
$$f(x)+\g\,  \hbox{dist}(x;X)$$
over $Y$.
}

\proof Denote $F(x)=f(x)+\g\,  \hbox{dist}(x;X)$. For a vector $x\in Y$, let $\hat x$ denote a vector of $X$ that is at minimum distance from $X$. If  $\g>L$, we have
$$F(x)=f(x)+\g \|x-\hat x\|=f(\hat x)+\big(f(x)-f(\hat x)\big)+\g \|x-\hat x\|\ge f(\hat x)+(\g-L) \|x-\hat x\|\ge F(\hat x),\qquad \forall\ x\in Y,$$
with strict inequality if $x\ne \hat x$; here the first inequality follows using the Lipschitz property of $f$ to write
$$f(x)-f(\hat x)\ge -L\|x-\hat x\|,$$
while the second inequality follows from the fact $f(\hat x)=F(\hat x)$. In words, the value of $F(x)$ is strictly reduced when we project an $x\in Y$ with $x\notin X$ onto $X$. 
 Hence the minima of $F$ over $Y$ can only lie within $X$, while $F=f$ within $X$. Thus all minima of $F$ over $Y$ must lie in $X$ and also minimize $f$ over $X$ (since $F=f$ on $X$). Conversely, all minima of $f$ over $X$ are also minima of $F$ over $X$ (since $F=f$ on $X$), and by the preceding inequality, they are also minima of $F$ over $Y$. \qed

We now provide a generalization for $m>1$.\footnote{\dag}{\ninepoint  In the original version of this report the assumption on existence of the scalar $\b$ in the proposition below was neglected, due to a faulty application of Prop.\ \propgenexact\ in its proof. This was noted in a paper by Kundu, Bach, and Bhattacharrya in Oct.\ 2017. If the sets $X_i$ are polyhedral this assumption is not necessary; this is Hoffman's lemma.}

\xdef\propgenexacto{\propn}\propnum\show{myproposition}

\texshopbox{\proposition{\propgenexacto:} 
Let $f:Y\mapsto \re$ be a function defined on a subset $Y$ of $\rn$, and let $X_i$, $i=1,\ldots,m$, be closed subsets of $Y$ with nonempty intersection.
Assume that $f$ is Lipschitz continuous over $Y$ with constant $L$, and that for some scalar $\b>0$, we have
$$\hbox{dist}(x;X_1\cap\cdots\cap X_m)\le \b\sum_{i=1}^m\hbox{dist}(x;X_i),\qquad \forall\ x\in Y.\xdef\regcondition{\lab}\eqnum\show{first}$$
Let $\g$ be a scalar with $\g>\b L$. Then the set of minima of $f$ over
$\cap_{i=1}^mX_i$ coincides with the set of minima of 
$$f(x)+\g  \sum_{i=1}^m\hbox{dist}(x;X_i)$$
over $Y$.
}

\proof The proof is similar to the proof of Prop.\ \propgenexact, using Eq.\ \regcondition\ to modify the main inequality. Denote $F(x)=f(x)+\g\,  \sum_{i=1}^m\hbox{dist}(x;X_i)$ and $X=X_1\cap\cdots\cap X_m$. 
For a vector $x\in Y$, let $\hat x_i$ denote a vector of $X_i$ that is at minimum distance from $x$, and let $\hat x$ denote a vector of $X$ that is at minimum distance from $x$. If  $\g>\b L$, we have
$$F(x)=f(x)+\g \sum_{i=1}^m\|x-\hat x_i\|\ge f(\hat x)+\big(f(x)-f(\hat x)\big)+{\g\over \b}\, \|x-\hat x\|\ge f(\hat x)+\lf ({\g\over \b}-L\ri) \|x-\hat x\|\ge F(\hat x),\qquad \forall\ x\in Y,$$
with strict inequality if $x\ne \hat x$. The proof now proceeds as in the proof of Prop.\ \propgenexact. \qed 

Regarding algorithmic solution, from Prop.\ \propgenexacto, it follows that we may consider in place of the original problem \iterprojprob\ the additive cost  problem \iterprojpen\ for which our algorithms apply. In particular, let us consider the algorithms \combincrsubgrt-\combincrsubgrbt, with $X=\rn$, which involve a proximal iteration on one of the functions $\g\,\hbox{dist}(x;X_i)$ followed by a subgradient iteration on $f$. A key fact here is that the proximal iteration
$$z_k=\arg\min_{x\in \rn}\left\{\g \,\hbox{dist}(x;X_{i_k})+{1\over 2\a_k}\|x-x_k\|^2\right\}\xdef\iterprojprox{\lab}\eqnum\show{first}$$
involves a projection on $X_{i_k}$ of $x_k$, followed by an interpolation.
This is shown in the following proposition.
 
\xdef\propiterproj{\propn}\propnum\show{myproposition}

\texshopbox{\proposition{\propiterproj:} Let  $z_k$ be the vector produced by the proximal iteration \iterprojprox.
If $x_k\in X_{i_k}$ then $z_k=x_k$, while if $x_k\notin X_{i_k}$, 
$$z_k=\cases{(1-\b_k)x_k+\b_k P_{X_{i_k}}(x_k) &if $\b_k <1$,\cr
P_{X_{i_k}}(x_k)&if $\b_k\ge 1$,\cr}\xdef\iterprojproxt{\lab}\eqnum\show{first}$$
where
$$\b_k={\a_k \g \over \hbox{dist}(x_{k};X_{i_k})}.$$
}

\proof The case $x_k\in X_{i_k}$ is evident, so assume that $x_k\notin X_{i_k}$. From the nature of the cost function in Eq.\ \iterprojprox\ we see that $z_k$ is a vector that lies in the line segment between $x_k$ and $P_{X_{i_k}}(x_k)$. Hence there are two possibilities: either
$$z_k=P_{X_{i_k}}(x_k),\xdef\caseone{\lab}\eqnum\show{first}$$
or $z_k\notin X_{i_k}$ in which case by setting to 0 the gradient at $z_k$ of the cost function in Eq.\ \iterprojprox\ yields
$$\g \,{z_k-P_{X_{i_k}}(z_k)\over \lf\|z_k-P_{X_{i_k}}(z_k)\ri\|}={1\over \a_k}(x_k-z_k).$$
This equation implies that $x_k$, $z_k$, and $P_{X_{i_k}}(z_k)$ lie on the same line, so that  $P_{X_{i_k}}(z_k)=P_{X_{i_k}}(x_k)$ and
$$z_k=x_k - {\a_k \g \over \hbox{dist}(x_{k};X_{i_k})}\big(x_k-P_{X_{i_k}}(x_k)\big)=(1-\b_k)x_k+\b_k P_{X_{i_k}}(x_k).\xdef\casetwo{\lab}\eqnum\show{first}$$
By calculating and comparing the value of the cost function  in Eq.\ \iterprojprox\ for each of the possibilities \caseone\ and \casetwo, we can verify that \casetwo\ gives a lower cost if and only if $\b_k < 1$. \qed

Let us now consider the problem
$$\eqalign{\hbox{\rm minimize}\quad &
\sum_{i=1}^m\big(f_i(x)+h_i(x)\big)\cr
\hbox{\rm subject to\ \ }
&x\in \cap_{i=1}^mX_i.\cr}$$
Based on the preceding analysis, we can convert this problem to the unconstrained minimization problem
$$\eqalign{\hbox{\rm minimize}\quad &
\sum_{i=1}^m\big(f_i(x)+h_i(x)+\g \hbox{dist}(x;X_i)\big)\cr
\hbox{\rm subject to\ \ }
&x\in \rn,\cr}$$
where $\g$ is sufficiently large. The algorithm \combincrsubgrbt, applied to this problem, yields the iteration
$$y_k=x_k-\a_k \suh_{i_k}(x_k),\qquad z_k=y_k-\a_k \suf_{i_k}(z_k),\qquad x_{k+1}=\cases{(1-\b_k)z_k+\b_k P_{X_{i_k}}(z_k) &if $\b_k <1$,\cr
P_{X_{i_k}}(z_k)&if $\b_k\ge 1$,\cr}$$
where
$$\b_k={\a_k \g \over \hbox{dist}(z_{k};X_{i_k})},$$
[cf.\ Eq.\ \iterprojproxt]. The index $i_k$ may be chosen either randomly or according to a cyclic rule.

Let us finally note another problem where our incremental methods apply:
$$\eqalign{\hbox{\rm minimize}\quad &
f(x)+c\sum_{j=1}^r\max\big\{0,g_j(x)\big\}\cr
\hbox{\rm subject to\ \ }
&x\in \cap_{i=1}^mX_i.\cr}$$
This type of problem is obtained by replacing convex inequality constraints of the form $g_j(x)\le0$ with the nondifferentiable penalty terms  $c\max\big\{0,g_j(x)\big\}$, where $c>0$ is a penalty parameter. Then a possible incremental method at each iteration, would either do a subgradient or proximal iteration on $f$, or select one of the violated constraints (if any) and perform a subgradient iteration on the corresponding function $g_j$, or select one of the sets $X_i$ and do an interpolated projection on it. Related methods may also be obtained when $f$ is replaced by a cost function of the form
$$\sum_{i=1}^m\big(f_i(x)+h_i(x)\big),$$
and the components $f_i$ are dealt with  a proximal iteration while  the components $h_i$ are dealt with  a subgradient iteration.

\vskip  -1.5pc
\section{Conclusions}
\vskip  -0.5mm
\mark{Conclusions}

\pn We have surveyed incremental algorithms, which can deal with many of the challenges  posed by large data sets in machine learning applications, as well as with the additive structure of many interesting problems, including those arising in the context of  duality. We have used a unified analytical framework that includes incremental proximal algorithms and their combinations with the more established incremental gradient and subgradient methods. This allows the  flexibility to separate the cost function into the parts that are conveniently handled by proximal iterations (e.g., in essentially closed form), and the remaining parts to be handled by subgradient iterations. We have outlined the convergence properties of these methods, and we have shown that our algorithms apply to some important problems that have been the focus of recent research. 

Much work remains to be done to apply and evaluate our methods within the broad context of potential applications. Let us mention some possibilities that may extend the range of applications of our approach, and are interesting subjects for further investigation: alternative proximal and projected subgradient iterations, involving  nonquadratic proximal terms and/or subgradient projections,  alternative stepsize rules, distributed asynchronous implementations along the lines of [NBB01], polyhedral approximation (bundle) variants of the proximal iterations in the spirit of [BeY09], and variants for methods with errors in the calculation of the subgradients along the lines of [NeB10].

\vskip  -1.5pc
\section{References}
\vskip  -2mm
\mark{References}

\def\refer{\vskip4pt\par\noindent}

\ninepoint

\ref[BCL03] Bauschke, H.\ H., Combettes, P.\ L., and Luke, D.\ R., 2003.\ ``Hybrid Projection-Reflection Method for
Phase Retrieval," Journal of the Optical Society of America, Vol.\ 20, pp.\ 1025-1034.

\ref[BCK06] Bauschke, H.\ H., Combettes, P.\ L., and Kruk, S.\ G., 2006.\ ``Extrapolation Algorithm for Affine-Convex Feasibility Problems," Numer.\ Algorithms, Vol.\ 41, pp.\ 239-274.
 
\ref[BHG08] Blatt, D., Hero, A.\ O., Gauchman, H., 2008.\ ``A Convergent Incremental Gradient Method with a Constant Step Size," SIAM J.\ Optimization, Vol.\ 18, pp.\ 29-51.

\ref [BMN01]
Ben-Tal, A., Margalit, T., and Nemirovski, A., 2001.\
``The Ordered Subsets Mirror Descent Optimization Method and its Use for 
the Positron Emission Tomography Reconstruction,'' in Inherently Parallel Algorithms
in Feasibility and Optimization and their Applications (D.\ Butnariu, Y.\ Censor,
and S.\ Reich, eds.), Elsevier, Amsterdam, Netherlands.

\ref [BMS99] Boltyanski, V., Martini, H., and Soltan, V., 1999.\ 
Geometric Methods and Optimization Problems, Kluwer,
Boston.

\ref[BNO03] Bertsekas, D.\ P., Nedi\'c, A., and Ozdaglar, A.\ E., 2003.\
Convex Analysis and Optimization,  Athena Scientific, Belmont, MA.

\ref[BPC10] Boyd, S., Parikh, N., Chu, E., Peleato, B., and Eckstein, J., 2010.\ ``Distributed Optimization and Statistical Learning via the Alternating Direction Method of Multipliers," working paper on line, Stanford, Univ.

\ref[Bau01] Bauschke, H.\ H., 2001.\ ``Projection Algorithms: Results and Open Problems," in Inherently Parallel Algorithms
in Feasibility and Optimization and their Applications (D.\ Butnariu, Y.\ Censor,
and S.\ Reich, eds.), Elsevier, Amsterdam, Netherlands.

\ref
[BeT89] Bertsekas, D.\ P., and Tsitsiklis, J.\ N., 1989.\ Parallel and
Distributed Computation: Numerical Methods, Prentice-Hall, Englewood Cliffs,
N.\ J.

\ref[BeT96] Bertsekas, D.\ P.,  and Tsitsiklis, J.\ N., 1996.\ Neuro-Dynamic Programming, Athena Scientific, Belmont, MA.

\ref[BeT00] Bertsekas, D.\ P.,  and Tsitsiklis, J.\ N., 2000.\ ``Gradient Convergence in Gradient Methods," SIAM J.\ Optimization, Vol.\ 10, pp.\ 627-642.

\ref[BeT09] Beck, A., and Teboulle, M., 2009.\ ``A Fast Iterative Shrinkage-Thresholding Algorithm for Linear
Inverse Problems," SIAM J.\ on Imaging Sciences, Vol.\ 2, pp.\ 183-202.

\ref[BeT10] Beck, A., and Teboulle, M., 2010.\ ``Gradient-Based Algorithms with Applications to Signal-Recovery Problems," in Convex
Optimization in Signal Processing and Communications (Y. Eldar and D. Palomar, eds.),
Cambridge University Press, pp.\ 42-88.

\ref[BeY09] Bertsekas, D.\ P.,  and Yu, H., 2009.\ ``A Unifying Polyhedral Approximation Framework for Convex Optimization," Lab.\ for Information and Decision Systems Report LIDS-P-2820, MIT; to appear in SIAM J. on Optimization.

\ref[Ber83] Bertsekas, D.\ P.,  1983.\ ``Distributed Asynchronous Computation of Fixed Points," Mathematical Programming, Vol.\ 27, pp.\ 107-120.

\ref[Ber96] Bertsekas, D.\ P.,  1996.\ ``Incremental Least Squares Methods and the Extended Kalman Filter," SIAM J.\
on Optimization, Vol.\ 6, pp.\ 807-822.

\ref [Ber97] Bertsekas, D.\ P., 
1997.\ ``A Hybrid Incremental Gradient Method for Least
Squares," SIAM J. on Optimization, Vol.\ 7,
pp.\ 913-926. 

\ref[Ber99] 
Bertsekas, D.\ P., 1999.\
Nonlinear Programming, 2nd edition, Athena Scientific, Belmont, MA.

\ref[Ber09] Bertsekas, D.\ P.,  2009.\ Convex Optimization Theory, Athena
Scientific, Belmont, MA; also, this book's on-line supplementary chapter on algorithms.

\ref[Ber10] Bertsekas, D.\ P.,  2010.\ ``Incremental Proximal Methods for Large Scale Convex Optimization," Lab.\ for Info.\ and Decision Systems Report
LIDS-P-2847, MIT, Cambridge, MA; to appear in Math.\ Programming J.

\ref[BiF07] Bioucas-Dias, J., and Figueiredo, M.\ A.\ T., 2007.\ ``A New TwIST: Two-Step Iterative Shrinkage/Thresholding Algorithms
for Image Restoration," IEEE Trans.\ Image Processing, Vol.\ 16, pp.\ 2992-3004.

\old{
\ref[BlH04] Blatt, D., and Hero, A., 2004.\ ``Distributed Maximum Likelihood for Sensor Networks," Proc.\
of the 2004 IEEE International Conference on Acoustics, Speech, and Signal Processing,
Montreal, Canada, pp.\ 929-932.
}

\ref[Bor08] Borkar, V.\ S., 2008.\ 
Stochastic Approximation: A Dynamical Systems Viewpoint, Cambridge Univ. Press.

\ref[Bot05] Bottou, L., 2005.\ ``SGD: Stochastic Gradient Descent," http://leon.bottou.org/projects/sgd.

\ref[CDL98] Chambolle, A., DeVore, R.\ A., Lee, N.\ Y., and Lucier, B.\ J., 1998.\ ``Nonlinear Wavelet Image Processing:
Variational Problems, Compression, and Noise Removal Through Wavelet Shrinkage," IEEE Trans.\ Image Processing, Vol.\ 7, pp.\ 319-335.

\ref[CeS08] Cegielski, A., and Suchocka, A., 2008.\ ``Relaxed Alternating Projection Methods," SIAM J.\ Optimization, Vol.\ 19, pp.\ 1093-1106.

\ref[CoW05] Combettes, P.\ L., and Wajs, V.\ R., 2005.\ ``Signal Recovery by Proximal Forward-Backward Splitting," Multiscale
Modeling and Simulation, Vol.\ 4, pp.\ 1168-1200.

\ref[DDM04] Daubechies, I., Defrise, M., and Mol, C. D., 2004.\ ``An Iterative Thresholding Algorithm for Linear Inverse
Problems with a Sparsity Constraint," Comm.\ Pure Appl.\ Math., Vol.\ 57, pp.\ 1413-1457.

\ref[DrH04] Drezner, Z., and Hamacher, H.\ W., 2004.\ Facility Location: Applications and Theory, Springer, N.\ Y.

\ref[DHS10] Duchi, J., Hazan, E., and Singer, Y., 2010.\ ``Adaptive Subgradient Methods for Online Learning and Stochastic Optimization," UC Berkeley EECS Technical Report 2010-24, to appear in J.\ of Machine Learning Research.

\ref[Dav76] Davidon, W.\ C., 1976.\ ``New Least Squares Algorithms,''
J.\ Optimization Theory and Applications, Vol.\ 18, pp.\ 187-197.

\ref[EMZ07] Elad, M., Matalon, B., and Zibulevsky, M., 2007.\ ``Coordinate and Subspace Optimization Methods for Linear Least Squares with Non-Quadratic Regularization," J.\ on Applied and Computational Harmonic Analysis, Vol.\ 23, pp.\ 346-367.

\ref[EcB92] Eckstein, J.,  and Bertsekas, D.\ P., 1992.\ ``On the Douglas-Rachford Splitting Method and the Proximal Point Algorithm for Maximal Monotone Operators," Math.\
Programming, Vol.\ 55, pp.\ 293-318.

\ref[Erm69] Ermoliev, Yu.\ M.,
``On the Stochastic Quasi-Gradient Method
and Stochastic Quasi-Feyer Sequences,'' 
Kibernetika, No.\ 2, 1969, pp.\ 73--83.

\ref[Erm76]
Ermoliev, Yu.\ M.,
Stochastic Programming Methods,
Nauka, Moscow, 1976.

\ref[FiN03] Figueiredo, M.\ A.\ T., and Nowak, R.\ D., 2003.\ ``An EM Algorithm for Wavelet-Based Image Restoration," IEEE
Trans.\ Image Processing, Vol.\ 12, pp.\ 906-916.

\ref[FNW07] Figueiredo, M.\ A.\ T., Nowak, R.\ D., and Wright, S.\ J., 2007.\ ``Gradient Projection for Sparse Reconstruction:
Application to Compressed Sensing and Other Inverse Problems," IEEE J.\ Sel.\ Topics in Signal Processing, Vol.\ 1, pp.\ 586-597.

\ref[GGM06] Gaudioso, M., Giallombardo, G., and Miglionico, G., 2006.\ ``An Incremental Method for Solving Convex Finite Min-Max Problems," Math.\ of Operations Research, Vol.\ 31,  pp.\ 173-187. 

\ref[GMS10] Goldfarb, D., Ma, S., and Scheinberg, K., 2010.\  ``Fast Alternating Linearization Methods for Minimizing the Sum of Two Convex Functions", Columbia Univ.\ report, on line.

\ref[GPR67] Gubin, L.\ G., Polyak, B.\ T., and Raik, E.\ V., 1967.\ ``The Method of Projection for Finding the Common
Point in Convex Sets," U.S.S.R. Comput.\ Math.\ Phys., Vol.\ 7, pp.\ 1Ð24 (English Translation).

\ref[GaM76] Gabay, D., and Mercier, B., 1979.\ ``A Dual Algorithm for the Solution of Nonlinear Variational Problems via Finite-Element Approximations," Comp.\ Math.\ Appl., Vol.\ 2, pp.\ 17-40.

\ref[Gab83] Gabay, D., 1983.\ ``Applications of the Method of Multipliers to Variational Inequalities," in M.\ Fortin and R.\ Glowinski, eds., Augmented Lagrangian Methods: Applications to the Solution of Boundary-Value Problems, North-Holland, Amsterdam.

\ref[GoM09] Goldfarb, D., and Ma, S., 2009.\ ``Fast Multiple Splitting Algorithms for Convex Optimization," Columbia Univ.\ report, on line.

\ref[Gri94]
Grippo, L., 1994.\
``A Class of Unconstrained Minimization Methods
for Neural Network Training,''
Optim.\ Methods and Software,
Vol.\ 4, pp.\ 135-150.

\ref[Gri00]
Grippo, L., 2000.\
``Convergent On-Line Algorithms for Supervised Learning in Neural Networks," IEEE Trans.\
Neural Networks, Vol.\ 11, pp.\ 1284-1299.

\ref[HeD09] Helou, E.\ S., and De Pierro, A.\ R., 2009.\ ``Incremental Subgradients for Constrained Convex Optimization: A Unified Framework and New Methods,"  SIAM J.\ on Optimization, Vol.\ 20, pp.\ 1547-1572.

\ref[JRJ09] Johansson, B., Rabi, M., and Johansson, M., 2009.\ ``A Randomized Incremental Subgradient Method for Distributed Optimization in Networked Systems," SIAM J.\ on Optimization, Vol.\ 20, pp.\ 1157-1170.

\ref[Kib80]
Kibardin, V.\ M., 1980.\
``Decomposition into Functions in the Minimization Problem,'' Automation and Remote Control,
Vol.\ 40, pp.\ 1311-1323.

\ref[Kiw04]
Kiwiel, K.\ C., 2004.\ 
``Convergence of Approximate and Incremental Subgradient Methods
for Convex Optimization,'' SIAM J.\ on Optimization, Vol.\ 14, pp.\ 807-840.

\ref[KuC78] Kushner, H.\ J., and Clark, D.\ S., 1978.\ Stochastic Approximation Methods for
Constrained and Unconstrained Systems, Springer-Verlag, N.\ Y.

\ref[KuY97] Kushner, H.\ J., and Yin, G., 1997.\ Stochastic Approximation Methods, Springer-Verlag, N.\ Y.

\ref[LMY08]
Lu, Z., Monteiro, R.\ D.\ C., and Yuan, M., 2008.\ 
``Convex Optimization Methods for Dimension
Reduction and Coefficient Estimation in Multivariate Linear Regression," 
Report, School of Industrial and Systems Engineering, 
Georgia Institute of Technology, Atlanta; appeared on line in Math.\ Programming J., 2010.

\ref[LeW10] Lee, S., and Wright, S.\ J., 2010.\ ``Sparse Nonlinear Support Vector Machines via Stochastic Approximation," Univ.\ of Wisconsin Report, submitted. 

\ref[LiM79] Lions, P.\ L., and Mercier, B., 1979.\ ``Splitting Algorithms for the Sum of Two Nonlinear Operators," SIAM J.\ on Numerical Analysis, Vol.\ 16, pp.\ 964-979.

\ref [Lit66] Litvakov, B.\ M., 1966.\ ``On an Iteration Method in the Problem of Approximating a Function from a Finite Number of Observations," Avtom.\ Telemech., No.\ 4, pp.\ 104-113.

\ref[Lju77] Ljung, L., 1977.\ ``Analysis of Recursive Stochastic 
Algorithms,''
IEEE Trans.\ on Automatic Control, Vol.\ 22, pp.\ 551-575.

\ref[LuT94] Luo, Z.\ Q., and Tseng, P., 1994.\ ``Analysis of an Approximate Gradient Projection Method with Applications to the Backpropagation Algorithm," Optimization Methods and Software,
Vol.\ 4, pp. 85-101. 

\ref[Luo91] Luo, Z.\ Q., 1991.\ ``On the Convergence of the LMS Algorithm with Adaptive Learning Rate for
Linear Feedforward Networks," Neural Computation, Vol.\ 3, pp.\ 226-245.

\ref[MYF03] Moriyama, H., Yamashita N., and Fukushima, M., 2003.\ ``The Incremental Gauss-Newton Algorithm with Adaptive Stepsize Rule," 
Computational Optimization and Applications,
Vol.\ 26, pp.\ 107-141.

\ref[MaS94]
Mangasarian, O.\ L., and Solodov, M.\ V., 1994.\ 
``Serial and Parallel Backpropagation Convergence
Via Nonmonotone Perturbed Minimization,''
Opt.\ Methods and Software,
Vol.\ 4, pp.\ 103-116.

\ref[Mar70]  Martinet, B., 
1970.\ ``Regularisation d' In\' equations Variationelles par
Approximations Successives," Revue Fran.\ d'Automatique 
et Infomatique Rech.\ Op\' erationelle, Vol.\ 4, pp.\ 154-159.

\refer[Mey07] Meyn, S., 2007.\ Control Techniques for Complex Networks, Cambridge University Press, N.\ Y.

\ref[NBB01]
Nedi\'c, A., Bertsekas, D.\ P., and Borkar, V., 2001.\
``Distributed Asynchronous Incremental Subgradient Methods,'' in Inherently Parallel Algorithms
in Feasibility and Optimization and their Applications (D.\ Butnariu, Y.\ Censor,
and S.\ Reich, eds.), Elsevier, Amsterdam, Netherlands.

\ref[NJL09] Nemirovski, A., Juditsky, A., Lan, G., and Shapiro, A., 2009.\
``Robust Stochastic Approximation Approach to
Stochastic Programming," SIAM Journal on
Optimization, Vol.\ 19, pp.\ 1574-1609.

\ref[NeB00]
Nedi\'c, A., and Bertsekas, D.\ P., 2000.\
``Convergence Rate of the Incremental Subgradient Algorithm,'' in 
Stochastic Optimization: Algorithms and
Applications, Eds., S.\ Uryasev and P.\ M.\ Pardalos,  
Kluwer Academic Publishers, pp.\ 263-304.

\ref[NeB01]
Nedi\'c, A., and Bertsekas, D.\ P., 2001.\
``Incremental Subgradient Methods for Nondifferentiable Optimization,''
SIAM J.\ on Optimization,
Vol.\ 12, 2001, pp.\ 109-138.

\ref[NeB10] Nedi\'c, A., and Bertsekas, D.\ P., 2010.\
``The Effect of Deterministic Noise in Subgradient Methods," Math.\ Programming, Ser.\  A, Vol.\ 125, pp.\ 75-99. 

\ref[NeO09] Nedi\'c, A., and Ozdaglar, A., 2009.\
``Distributed Subgradient Methods for Multi-Agent Optimization," IEEE Trans.\ on Aut.\ Control, Vol.\ 54, pp.\ 48-61.

\ref[Ned10] Nedi\'c, A., 2010.\
``Random Projection Algorithms for Convex Minimization Problems," Univ.\ of Illinois Report; appear in Math.\ Programming Journal.

\ref[Nes83]
Nesterov, Y., 1983.\ 
``A Method for Unconstrained Convex Minimization Problem with the Rate of Convergence $O(1/k^2)$," Doklady AN SSSR 269, pp.\ 543-547; translated as Soviet Math. Dokl.

\ref[Nes04] Nesterov, Y., 2004.\  Introductory Lectures on Convex Optimization, Kluwer Academic Publisher,
Dordrecht, The Netherlands.

\ref[Nes05] Nesterov, Y., 2005.\ ``Smooth Minimization of Nonsmooth Functions," Math.\ Programming, Vol.\ 103 pp.\ 127-152.

\ref[Nev75] Neveu, J., 1975.\ Discrete Parameter Martingales, North-Holland, Amsterdam, The Netherlands.

\ref[PKP09] Predd, J.\ B., Kulkarni, S.\ R., and Poor, H.\ V., 2009.\ ``A Collaborative Training Algorithm for Distributed Learning," IEEE Transactions on Information Theory,
Vol.\ 55,  pp.\ 1856-1871.

\ref[Pas79] Passty, G.\ B., 1979.\ ``Ergodic Convergence to a Zero of the Sum of Monotone Operators in Hilbert Space," J.\ Math.\
Anal.\ Appl., Vol.\ 72, pp.\ 383-390.

\ref[Pfl96] Pflug, G., 1996.\ Optimization of Stochastic Models. The Interface Between Simulation and
Optimization, Kluwer, Boston.

\ref[PoT73] Polyak, B.\ T., and Tsypkin, Y.\ Z., 1973.\ ``Pseudogradient 
Adaptation and Training Algorithms,'' Automation and Remote Control,
Vol.\ 12, pp.\ 83-94.

\ref[Pol64] Poljak, B.\ T., 1964.\  ``Some Methods of Speeding up the Convergence
of Iteration Methods," Z.\ Vy\u Cisl.\ Mat.\ i Mat.\ Fiz., Vol.\ 4, pp.\ 1-17.

\ref[Pol87]
Polyak, B.\ T., 1987.\ Introduction to Optimization, Optimization Software Inc., N.\ Y.

\ref[Pol78]
Polyak, B.\ T., 1978.\ 
``Nonlinear Programming Methods in the Presence of Noise,''
Math.\ Programming, Vol.\ 14, pp.\ 87--97.

\ref[Pol87]
Polyak, B.\ T., 1987.\
Introduction to Optimization,
Optimization Software Inc., N.\ Y.

\ref[RNV09] Ram, S.\ S., Nedi\'c, A., and Veeravalli, V.\ V., 2009.\  ``Incremental Stochastic Subgradient Algorithms for Convex Optimization," SIAM Journal on Optimization, Vol.\ 20, pp.\ 691-717.

\ref[RNV10] Ram, S.\ S., Nedi\'c, A., and Veeravalli, V.\ V., 2010.\  ``Distributed Stochastic Subgradient Projection Algorithms for Convex Optimization," Journal of Optimization Theory and Applications, Vol.\ 147, pp.\ 516-545.

\ref[RaN04]
Rabbat, M.\ G., and Nowak, R.\ D., 2004.\ ``Distributed Optimization in Sensor Networks," in Proc.\ Inf.\ Processing Sensor Networks, Berkeley, CA, pp.\ 20-27.

\ref[RaN05]
Rabbat M.\ G., and Nowak R.\ D., 2005.\ ``Quantized Incremental Algorithms for Distributed Optimization,''  IEEE Journal on Select Areas in Communications,  Vol.\ 23, pp.\ 798-808. 

\ref [Roc70] Rockafellar, R.\ T., 1970.\ Convex Analysis, Princeton University
Press, Princeton, NJ.

\ref [Roc76]  Rockafellar, R.\ T., 1976.\ ``Monotone Operators and the Proximal Point
Algorithm," SIAM Journal on Control and Optimization, Vol.\ 14, pp.\ 877-898.

\ref[SSS07] Shalev-Shwartz, S., Singer, Y., Srebro, N., and  Cotter, A., 2007.\ ``Pegasos:
Primal Estimated Subgradient Solver for SVM," in
ICML Õ07, New York, N.\ Y., pp.\ 807-814.
 
\ref[SoZ98] Solodov, M.\ V., and Zavriev, S.\ K., 1998.\ ``Error Stability Properties of Generalized Gradient-Type Algorithms,'' J.\ Opt.\ Theory and Appl., Vol.\ 98, pp.\ 663-680.

\ref[Sol98] Solodov, M.\ V., 1998.\ ``Incremental Gradient Algorithms with Stepsizes Bounded Away from Zero," Comput.\ Opt.\ Appl., Vol.\ 11, pp.\ 28-35.

\ref[Spi85] Spingarn, J.\ E., 1985.\ ``Applications of the Method of Partial Inverses to Convex Programming: Decomposition," Math.\ Programming, Vol.\ 32, pp.\ 199-223.

\ref[TBA86] Tsitsiklis, J.\ N.,  Bertsekas, D.\ P., and Athans, M., 1986.\ ``Distributed Asynchronous
Deterministic and Stochastic Gradient Optimization Algorithms," IEEE Trans. Automatic Control, Vol.\ AC-31, pp.\ 803-812.

\ref[Tse98] Tseng, P., 1998.\ ``An Incremental Gradient(-Projection) Method with Momentum Term and Adaptive Stepsize Rule," SIAM J.\ on Optimization, Vol.\ 8, pp.\ 506-531.

\ref[Tse08] Tseng, P., 2008.\ ``On Accelerated Proximal Gradient
Methods for 
Convex-Concave Optimization," Report, Math.\ Dept., Univ.\ of Washington.

\ref[VoU07] Vonesch, C., and Unser, M., 2007.\ ``Fast Iterative Thresholding Algorithm for Wavelet-Regularized Deconvolution,"
in Proc.\ SPIE Optics and Photonics 2007 Conference on Mathematical Methods: Wavelet
XII, Vol. 6701, San Diego, CA, pp.\ 1-5.

\ref[WNF08] Wright, S.\ J., Nowak, R.\ D., and Figueiredo, M.\ A.\ T., 2008.\ ``Sparse Reconstruction by Separable Approximation,"
in Proceedings of the IEEE International Conference on Acoustics, Speech and Signal Processing
(ICASSP 2008), pp.\ 3373-3376.

\ref[WiH60] Widrow, B., and Hoff, M.\ E., 1960.\ ``Adaptive Switching Circuits," Institute of Radio Engineers, Western Electronic Show and Convention, Convention Record, Part 4, 
pp.\ 96-104.

\end